\documentclass[%
 aip,
 amsmath,amssymb,
 reprint,%
]{revtex4-1}
\usepackage{graphicx}
\usepackage{dcolumn}
\usepackage{bm}

\usepackage[utf8]{inputenc}
\usepackage[T1]{fontenc}
\usepackage{mathptmx}
\usepackage{etoolbox}
\usepackage{bm}
\usepackage[usenames,dvipsnames]{xcolor}
\definecolor{amber}{rgb}{1.0, 0.75, 0.0}
\definecolor{bulgarianrose}{rgb}{0.28, 0.02, 0.03}
\definecolor{maroon(html/css)}{rgb}{0.5, 0.0, 0.0}
\definecolor{violet}{rgb}{0.56, 0.0, 1.0}
\makeatletter
\def\@email#1#2{%
 \endgroup
 \patchcmd{\titleblock@produce}
  {\frontmatter@RRAPformat}
  {\frontmatter@RRAPformat{\produce@RRAP{*#1\href{mailto:#2}{#2}}}\frontmatter@RRAPformat}
  {}{}
}%
\makeatother
\usepackage[version=3]{mhchem} 
\usepackage[T1]{fontenc}       
\usepackage{bm}
\usepackage{amsmath}
\usepackage{amssymb}
\usepackage{amsfonts}
\usepackage{color,soul}
\usepackage[utf8]{inputenc}
\usepackage{mathrsfs}
\usepackage{xcolor}
\usepackage{caption}
\usepackage{subcaption}

\DeclareUnicodeCharacter{0308}{\diaeresis}
\DeclareRobustCommand\diaeresis{%
\"}
\usepackage{comment}

\newcommand*{\pvec}[1]{\vec{#1}\mkern2mu\vphantom{#1}'}
\newcommand*{\abs}[1]{\left|#1\right|}

\begin{document}

\preprint{AIP/123-QED}

\title{Characterization of Excited States in Time-Dependent Density Functional Theory Using Localized Molecular Orbitals}
\author{Souloke Sen}
 \affiliation{Division of Theoretical Chemistry, Faculty of Sciences, Vrije Universiteit Amsterdam, De Boelelaan 1083, 1081 HV Amsterdam, The Netherlands}
\author{Bruno Senjean}
\affiliation{ICGM, Université de Montpellier, CNRS, ENSCM, Montpellier, France}
\author{Lucas Visscher}
\email{l.visscher@vu.nl}
\affiliation{Division of Theoretical Chemistry, Faculty of Sciences, Vrije Universiteit Amsterdam, De Boelelaan 1083, 1081 HV Amsterdam, The Netherlands}

\date{\today}

\begin{abstract}
Localized molecular orbitals are often used for the analysis of chemical bonds, but they can also serve to efficiently and comprehensibly compute linear response properties. 
While conventional canonical molecular orbitals provide an adequate basis for the treatment of excited states, a chemically meaningful identification of the different excited-state processes is difficult within such a delocalized orbital basis. 
In this work, starting from an initial set of supermolecular canonical molecular orbitals, we provide a simple one-step top-down embedding procedure for generating a set of orbitals which are localized in terms of the supermolecule, but delocalized over each subsystem composing the supermolecule. 
Using an orbital partitioning scheme based on such sets of localized orbitals, we further present a procedure for the construction of local excitations and charge-transfer states within the linear response framework of time-dependent density functional theory (TDDFT). 
This procedure provides direct access to approximate diabatic excitation energies and, under the Tamm--Dancoff approximation, also their corresponding electronic couplings -- quantities that are of primary importance in modelling energy transfer processes in complex biological systems.
Our approach is compared with a recently developed diabatization procedure based on subsystem TDDFT using projection operators, which leads to a similar set of working equations. 
Although both of these methods differ in the general localization strategies adopted and the type of basis functions (Slaters vs. Gaussians) employed, an overall decent agreement is obtained.
\end{abstract}

\maketitle

\section{Introduction}

Time-dependent density functional theory \cite{casida1995time,petersilka1996excitation,maitra2013successes,marques2004time,van1999implementation} (TDDFT) based on the Kohn--Sham \cite{kohn1965self} (KS) ground-state determinant is a popular method for the calculation of low-lying excited states of medium to large-sized molecules, owing to its favourable accuracy to cost ratio.\cite{marques2004time,van1999implementation,dreuw2005single,casida2009time,liu2011analytical}
After the advent of range-separated density functionals that are better suited for describing charge-transfer excitations,\cite{sarkar2021benchmarking,tawada2004long,iikura2001long,yanai2004new,stein2009reliable,laurent2013td,li2014comparison,moore2015charge} TDDFT has become the dominant approach for the study of photo-induced processes in biochromophoric systems. \cite{dreuw2005single,wormit2009development,wanko2005calculating,jacquemin2007assessment,jacquemin2008td,curutchet2017quantum} 
In such processes, excitation energy transfer (EET) and photo-induced electron transfer (ET) pathways play a crucial role.\cite{konig2012quantum,sariciftci1992photoinduced}
A prominent application is found in the study of antenna complexes of photosynthetic systems, responsible for the light-harvesting in higher plants.\cite{blankenship2002molecular,segatta2019quantum}
To understand how the initial energy transfer and subsequent charge separation happen, phenomenological models employ a simplified diabatic picture\cite{novoderezhkin2010physical,van2001understanding,novoderezhkin2007mixing,novoderezhkin2005pathways}
such as the Frenkel--Davydov exciton theory, in which the excited-state wavefunction is considered to be a superposition of local excitations (LEs)\cite{frenkel1931transformation,davydov1964theory} and for which several extensions have been proposed to better describe short-range interactions.\cite{abramavicius2009coherent, morrison2014ab,ma2013calculating,ma2012new,tretiak1999excitonic,li2017ab}
To connect TDDFT results to these models, it is necessary to express the supermolecular wave function in terms of localized molecular orbitals (LMOs) and to distinguish between LEs and charge-transfer (CT) excitations. 
Within this local or (approximate) diabatic picture,\cite{kurihara1997calculations,bouman1979optical,liu2015general,accomasso2019diabatization,cimiraglia1985quasi} ET and EET processes can be analyzed in a conceptually clear manner.\cite{liu2015general,li2017ab,accomasso2019diabatization,nottoli2018role,tolle2020electronic}

There are two dominant strategies
to obtain LMOs and LEs in a supermolecular system.\cite{dreuw2005single} 
One may either (1) \textit{a priori} partition the supermolecular system
into several subsystems and obtain LMOs from individual calculations on each subsystem,
or (2) partition the system \textit{after} a set of supermolecular orbitals is obtained. 

The first class of approaches to obtain all relevant supermolecular states in a diabatic picture belongs to the extensive family of embedding techniques.\cite{khait2010embedding,sun2016quantum,wesolowski2008embedding,wesolowski2014embedding,wesolowski2015frozen,daday2013state,hofener2012molecular,wen2019absolutely,chulhai2016external,goodpaster2011embedded,libisch2014embedded}
In particular, subsystem DFT,\cite{senatore1986,cortona1991self,wesolowski1993frozen,jacob2014subsystem,wesolowski2015frozen,tamukong2014density,chulhai2015frozen,tamukong2017accurate,kiewisch2008topological} along with its time-dependent extension in the linear response regime, has long been an efficient workhorse for DFT-in-DFT embedding\cite{casida2004generalization,neugebauer2007couplings,konig2013direct,neugebauer2009subsystem,konig2012quantum,neugebauer2008photophysical} where different levels of approximations can be used for different subsystems.\cite{tolle2022seamless,dresselhaus2015part} 
In the second class of approaches, LMOs are obtained by localizing the supermolecular canonical molecular orbitals (CMOs).\cite{foster1960canonical,edmiston1963localized,pipek1989fast,knizia2013intrinsic}
The resulting LMOs can then be used as a starting point for embedding calculations,\cite{lee2019projection,bennie2017pushing,hegely2016exact,tolle2019inter,henderson2006embedding} and/or to directly construct (approximate) diabatic states.\cite{subotnik2008constructing,subotnik2009initial,pavanello2011linking,tolle2020electronic} As shown by the Neugebauer group, subsystem TDDFT using projection-based embedding techniques can exactly reproduce supermolecular results and capture the complete range of LE/CT interactions.\cite{tolle2019inter,tolle2019exact,tolle2020electronic,scholz2020analysis,tolle2022seamless}

An elaborate discussion of the relations between these two approaches to localization of orbitals and approximate diabatization of electronic states is beyond the scope of this work, and we refer the reader to Refs.~\citenum{tolle2022seamless} and ~\citenum{bensberg2019automatic} for more details on these methods. 
To complete this brief list of relevant related methods, we also note the multistate fragmentation excitation difference-fragment charge difference (multistate FED-FCD) method that can also be used to build approximate diabatic states starting from supermolecular CMOs.\cite{voityuk2002fragment,hsu2008characterization,yang2013multi,nottoli2018role,cupellini2020charge,cupellini2018coupling}
By exploiting the spatial locality of LMOs, computationally efficient implementations of TDDFT can be achieved, as shown by Wu and co-workers who demonstrated linear scaling for TDDFT in combination with ``fragment LMOs''.\cite{wu2011linear} 
Wu and co-workers also nicely summarized the choices that need to be made in terms of different kinds of locality. 
Supermolecular CMOs are ``local in energy and delocalized in space'' while atomic orbitals (AOs) are ``local in space but delocalized in energy''. 
Both types of locality can furthermore be exploited by partially re-canonicalizing the orbital space.\cite{miura2009ab}
Such a partial re-canonicalization step is useful to improve the convergence of the iterative solution of the TDDFT equations as well as for interpretation purposes, as the orbital energies resulting from the re-canonicalization can be readily interpreted as perturbed subsystem energies.\\

In this work, we build on these previous developments and define a simple and computationally efficient strategy involving a one-step localization procedure to treat LE and CT states. 
Our approach can be regarded as an extension of our recent work on intrinsic localized fragment (molecular or atomic) orbitals, \cite{senjean2021generalization} by adding a re-canonicalization procedure.
This step results in a blocked Fock matrix that is similar to the one obtained with the embedding procedure by T{\"o}lle {\it et al.}\cite{tolle2020electronic}
Since our subsystem orbitals are orthogonal by construction, our procedure does, however, not require the construction of any projection operators.\cite{bensberg2019automatic,khait2010embedding,khait2012orthogonality}
The procedure allows for a partition of the full orbital space into LE and CT subspaces that are subsequently utilized in a TDDFT calculation to generate approximate diabatic states.
Using the Tamm--Dancoff approximation (TDA),\cite{tamm1945,*dancoff1950non} explicit expressions for the electronic couplings can be derived between these states.
Since our procedure is similar to projection-based subsystem DFT/TDDFT, we compare our result to Ref.~\citenum{tolle2020electronic}.
Note that the general framework presented here is particularly
appealing as it can easily be extended to any excited-state method.\\

The paper is organized as follows. 
In the first section, we provide the TDDFT working equations and then outline the construction of the re-canonicalized LMOs, followed by their application as a reference basis in the calculation of excitation energies and electronic couplings in the TDDFT framework. 
In the next section, a proof of principle application of our method on small model systems is provided, followed by an illustrative example of a chlorophyll dimer in which we also consider the use of additional tight-binding approximations. 
Conclusions and perspectives are then given in the final section.

\section{Theory}

\subsection{Time Dependent Density Functional Theory}

In the conventional formulation of TDDFT with real spatial orbitals in the adiabatic approximation,\cite{zangwill1980resonant,gross1985local,bauernschmitt1996treatment} excitation energies are determined by solving the non-Hermitian eigenvalue problem,
\begin{equation}
\begin{pmatrix}
 \textbf{A} & \textbf{B} \\ 
 -\textbf{B} &  -\textbf{A} \\
\end{pmatrix} 
\begin{pmatrix} 
\textbf{X} \\
\textbf{Y}\\
\end{pmatrix} = \mathbf{\omega}
\begin{pmatrix} 
\textbf{X} \\
\textbf{Y}\\
\end{pmatrix},  \label{TDDFT}
\end{equation} 
where the matrix elements of the $\mathbf{A}$ and $\mathbf{B}$ matrices are written as
\begin{equation}
A_{ia,jb} = \delta_{ij}F_{ab} - \delta_{ab}F_{ij} + K_{ia,jb}, \label{A}
\end{equation}
and
\begin{equation}
B_{ia,jb} = K_{ia,bj}.\label{B}
\end{equation}
In the above equation, indices $i,j$ denote occupied orbitals and $a,b$ denote virtual orbitals. 
$F_{pq}$ is an element of the KS Fock matrix, and the coupling matrix $\textbf{K}$ is defined as
\begin{align}
\label{K_definition}
K_{ia,jb} &= \int \mathrm{d}^3 \vec r \int \mathrm{d}^3 \pvec r \phi_i(\vec r) \phi_a(\vec r) f_\mathrm{Hxc}[\rho](\vec r, \pvec r) \; \phi_j(\pvec r) \phi_b(\pvec r) \nonumber\\ 
&-c_X\int \mathrm{d}^3 \vec r \int \mathrm{d}^3 \pvec r \phi_i(\vec r) \phi_j(\vec r) \frac{1}{\abs{\vec r - \pvec r}}\; \phi_a(\pvec r) \phi_b(\pvec r),
\end{align}
with the kernel
\begin{equation}\label{eq:Hxc_Kernel}
f_\mathrm{Hxc}[\rho^\mathrm{GS}](\vec r, \pvec r) = \frac{1}{\abs{\vec r - \pvec r}} + (1-c_X) \frac{\delta^2 E_\mathrm{xc}[\rho]}{\delta \rho(\vec r) \delta \rho(\pvec r)} \Big|_{\rho^\mathrm{GS}}
\end{equation}
consisting of the Coulomb term and the second derivative of the exchange-correlation energy functional $E_\mathrm{xc}[\rho]$. 
The second line of (\ref{K_definition}) allows for use of hybrid DFT approaches that mix in a fraction $c_X$ of exact exchange. 

In the CMO basis, the KS Fock matrix
$F_{pq}^{CMO} = \delta_{pq} \epsilon_{p}$
is diagonal
so that different excitation amplitudes $X_{ia}$ and $X_{jb}$ are only coupled by the coupling matrix $\textbf{K}$, i.e.,
\begin{equation}
A^{CMO}_{ia,jb} = \delta_{ij}\delta_{ab}(\epsilon_{a} - \epsilon_{i}) + K_{ia,jb},\label{A_CMO}
\end{equation}
where $\epsilon_p$ denotes the energy of orbital $p$.
The difference between canonical and non-canonical orbitals does not affect the $\mathbf{B}$ matrix that describes the coupling between excitations and de-excitations. 
For the sake of simplicity, we will employ the TDA in the following and include excitations only. 
The equation then reads
\begin{equation}
\textbf{A}\textbf{X} = \mathbf{\omega}\textbf{X},
\label{TDA}
\end{equation}
where the lowest eigenvalues and eigenvectors can be obtained by iterative diagonalization techniques.
In this work, we first partition the matrix $\textbf{A}$ into subsystems to be treated separately.

\subsection{Construction of Re-canonicalized Intrinsic Localized Molecular Orbitals} \label{constrct_RILMO}
In a previous work\cite{senjean2021generalization} we introduced the concept of (polarized) intrinsic fragment orbitals (IFOs), a minimal basis in which the supermolecular KS determinant can be expressed exactly. These are constructed from a basis $\mathit{B_{2}}$ of reference fragment orbitals (RFOs) that are usually selected as the core and valence orbitals of each fragment. 
After expressing the supermolecular CMO basis $\mathit{B_{1}}$ in terms of these IFOs, both the occupied and virtual valence spaces can be localized to generate a final set of so-called intrinsic localized molecular orbitals (ILMOs), using for instance Pipek--Mezey (PM) localization.\cite{pipek1989fast} 
The IFOs and ILMOs span the occupied space exactly but contain only $N_{B_2}-N_{occ}$ virtual orbitals. Such a truncation of the virtual space is undesirable for the current application to TDDFT, in which we want to retain a sufficient number of virtual orbitals to describe the electronically excited states accurately. 
A straightforward possibility is to increase the size of the RFO basis ($N_{B_2}$) by selecting more orbitals per fragment, but this will also affect the definition of the valence virtual orbitals that are convenient for interpretation purposes. 
Another possibility is to keep the minimal valence space as it is, but add a selected number of localized virtual orbitals from the complementary virtual space that is not spanned by the valence virtual orbitals. For this purpose, numerous methods have been developed\cite{liu2014localization}, which either explicitly rotate these virtual orbitals to increase their localization\cite{edmiston1963localized,pipek1989fast} or apply projections of the complementary virtual space onto a predefined set of ``proto-hard'' atomic virtual orbitals\cite{subotnik2005}.  

In this work, we explore an approach in which the localization of the complementary virtual space that consists of so-called ``hard virtual'' (hv) orbitals is carried out with the same algorithm as is used for the occupied space. This allows us to work with atomic or molecular fragments, or with a combination of both. Another advantage of our approach is that there is no bias for a particular fragment in defining the initial localization, the procedure can work with an arbitrary number of fragments and allocation of orbitals that are bonding two (or more) fragments together is done after the localization procedure. 
The key idea is to separate the supermolecular basis into four subspaces: $\mathit{B_{1}}=\mathit{B_{1c}}\cup\mathit{B_{1m}}\cup\mathit{B_{1v}}\cup\mathit{B_{1d}}$, with $\mathit{B_{1c}}$ indicating core orbitals, $\mathit{B_{1m}}$ indicating the minimal valence basis (comprising both occupied and virtual orbitals), $\mathit{B_{1v}}$ indicating energetically low-lying virtual orbitals, and $\mathit{B_{1d}}$ indicating energetically high-lying virtual orbitals. 
The core orbitals in space $\mathit{B_{1c}}$ are straightforwardly identified from the orbital energies of the supermolecular calculation and are removed from the orbital set before the localization procedures commences, i.e. we therein use space $\mathit{B_{1}}^\prime=\mathit{B_{1m}}\cup\mathit{B_{1v}}\cup\mathit{B_{1d}}$. 
This is a simple extension of our original implementation in which the (usually already very local) core orbitals were also localized. This is now an optional step with as advantage that the core-valence separation is kept exactly as it was in the CMO basis. For the RFOs that comprise reference basis $\mathit{B_{2}}$ we similarly partition the relevant part of the orbital space on each fragment as $\mathit{B_{2}}=\mathit{B_{2m}}\cup\mathit{B_{2v}}$. 
The procedure is then as follows: 

\begin{enumerate}
    \item Remove core orbitals from $\mathit{B_{1}}$ (basis $\mathit{B_{1}}^\prime$).
    \item Define valence RFOs (basis $\mathit{B_{2m}}$). 
    \item Compute valence IFOs (basis $\mathit{B_{1m}}$).
    \item Localize valence orbitals expressed in the IFO basis to obtain valence ILMOs.
    \item Define the complementary virtual space $\mathit{B_{1v}}\cup\mathit{B_{1d}}$.
    \item Re-canonicalize $\mathit{B_{1v}}\cup\mathit{B_{1d}}$ to obtain semicanonical supermolecular virtual orbitals  with effective energies $\epsilon^\prime$.
    \item Select $N_v$ orbitals with lowest values of $\epsilon^\prime$ (basis $\mathit{B_{1v}}$).
    \item Define hv RFOs (basis $\mathit{B_{2v}}$).
    \item Compute hv IFOs (basis $\mathit{B_{1v}}$).
    \item Localize the hv orbitals expressed in the IFO basis to obtain hv ILMOs.
\end{enumerate}

Steps 2 through 4 are identical to the procedure described in our previous work, reference \citenum{senjean2021generalization}, to which we refer for details about the algorithm and definitions of the valence RFOs, IFOs and ILMOs. Step 5 can also be done before step 4 (as it does not require information about the ILMOs) and is easily implemented by taking the eigenvectors corresponding to the zero eigenvalues of the singular value decomposition (Eq. 15 of reference \citenum{senjean2021generalization}) to define the complementary virtual space $\mathit{B_{1v}}\cup\mathit{B_{1d}}$. Steps 9 and 10 are carried out with exactly the same algorithm as steps 3 and 4 and are implemented by simply changing the arguments to the subroutines that perform these operations. The number of desired virtual orbitals $N_v$ thereby takes the place of the number $N_{occ}$ that is used when forming the occupied ILMOs. The advantage of this scheme is that mixing-in of high-energy virtual orbitals into the localized orbital set is strictly prohibited as rotations are only carried out between the lowest $N_v$ virtual orbitals. In this way we can effectively combine energy selection and orbital localization.

\begin{figure*}
\centering
\includegraphics[width=1\textwidth]{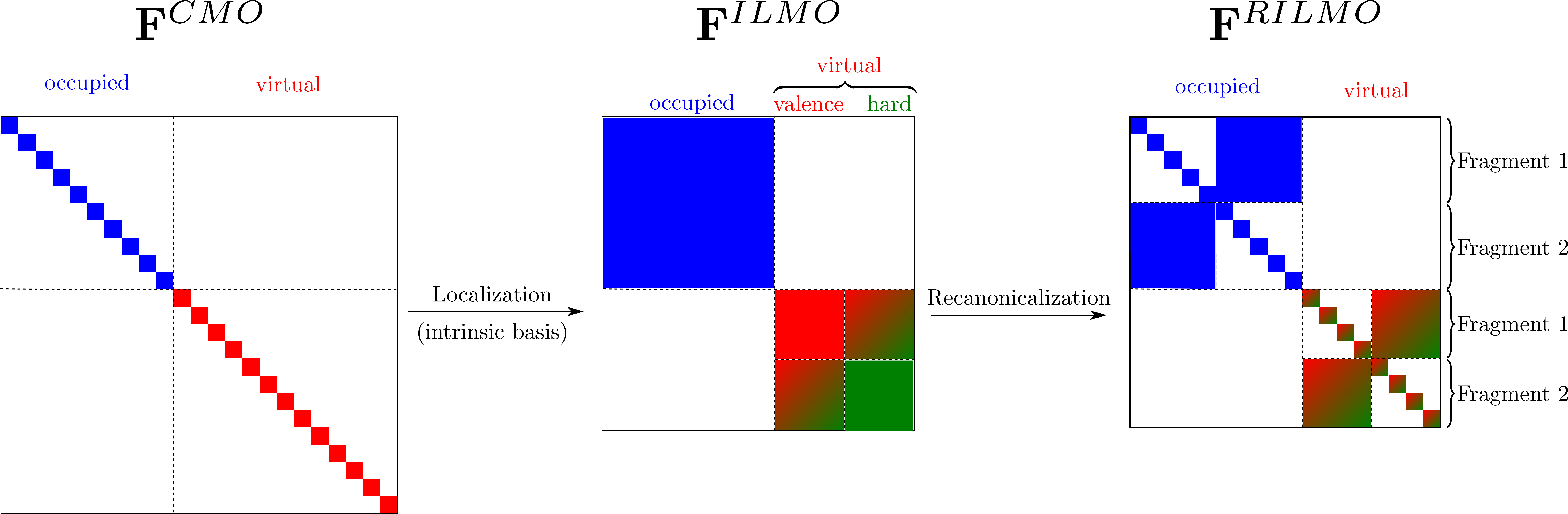}
\caption{Representation of the Fock matrix in different bases. Colored blocks represent non-zero matrix elements, where the smaller squares are single matrix elements.}\label{Fock}
\end{figure*}

Turning back towards our current application to partitioned TDDFT, we note that all localization transformations are unitary and do not mix the occupied and virtual orbitals that are obtained in the supermolecular calculation. 
They do therefore not change the structure of the TDDFT (Eq.\ref{TDDFT}) and TDA (Eq.\ref{TDA}) equations. 
The main difference compared to the CMO basis is the appearance of off-diagonal elements in the ILMO basis. 
In our previous work we observed that the diagonal elements of the transformed Fock matrix can provide an effective energy, but this pragmatic approach is not invariant for rotations that do not affect the localization extent (i.e. rotations between orbitals on the same fragment). It is therefore not well-suited for TDDFT where it is better to have a diagonally dominant $\textbf{A}$ matrix. In the current work, we therefore introduce a final re-canonicalization\cite{miura2009ab} of the occupied and virtual orbitals of each fragment that restores the diagonal dominance and also provides an unambiguous definition of the effective orbital energies.
To perform this re-canonicalization, we first sort the ILMO basis such that each orbital is assigned to the fragment on which it has its major contribution, according to Mulliken population analysis.\cite{mulliken1955electronic} This automatic assignment can optionally be fine-tuned for typical embedding cases in which dangling bonds should always be assigned to the fragment of primary interest. For brevity of notation, we assume from now on a partitioning of the supermolecule into two fragments $I$ and $J$, each with an integer number of occupied and virtual orbitals. 
After transformation of the supermolecular Fock matrix to this sorted ILMO basis, the subblocks of this Fock matrix, each in a basis of either an occupied or a virtual set of ILMOs, are diagonalized separately as shown in Fig.~\ref{Fock}.
The resulting eigenvectors form a unitary matrix $\mathbf{U}^{RILMO}$ that transforms the ILMO basis to the final re-canonicalized ILMO (RILMO) basis.
The Fock matrix in RILMO basis is partly diagonal,
\begin{eqnarray}
F^{RILMO}_{p\in I,q\in I} = \tilde{\epsilon}_p \delta_{pq}, \label{Flocal}\\
F^{RILMO}_{i\in I,a\in J} = 0,\\
F^{RILMO}_{i\in I,j\in J \neq I} \neq 0, \label{FCT1}\\
F^{RILMO}_{a\in I,b\in J \neq I} \neq 0 \label{FCT2},
\end{eqnarray}
with $\tilde{\epsilon}_p$ the eigenvalues obtained in this final re-canonicalization step.
These eigenvalues can readily be interpreted as orbital energies of the fragments, perturbed by the interaction with the other fragment(s). 
By choosing chemically relevant fragments, it is possible to identify $\pi$ and $\pi^{*}$ orbitals of conjugated aromatic systems that are delocalized over a single fragment only.

Given the similarity in structure of the Fock matrix in the RILMO basis with those arising in projection-based embedding, our re-canonicalization procedure can be seen as a one-step embedding where the resulting RILMOs are comparable to a set of polarized orthonormal subsystem orbitals resulting from a top-down embedding scheme.  \cite{tolle2020electronic,scholz2020analysis,tolle2019inter,tolle2019exact,chulhai2016external,bensberg2019automatic} 
The main advantage of the current procedure is its simplicity and efficiency: the generation of ILMOs is done via an iterative Jacobi algorithm that typically shows a rapid convergence, whereas the diagonalizations needed for the re-canonicalizations are done in subspaces of the full CMO space. The whole procedure therefore takes only a fraction of the time needed for the supermolecular DFT calculation and is carried out by a dedicated program requiring only the definition of the supermolecular and fragment orbitals and the atomic basis in which these are expressed.

Within the RILMO basis, characterization of LE and CT excitations for TDDFT calculations is easy. 
Choosing again two subsystems, 1 and 2, and labeling the excitations as either local ($L_1$ and $L_2$) or as electron transfer from system 1 to 2 ($CT_1$) or vice versa ($CT_2$) we obtain the same structure for the TDA equation (\ref{TDA}) as found in projection-based embedding \cite{tolle2019exact,tolle2019inter,tolle2020electronic,chulhai2016external}, but without explicit use of projectors.
\begin{widetext}
\begin{equation}
\begin{pmatrix}
 \mathbf{A}^{L_{1}} & 
 \mathbf{A}^{L_{1}/L_{2}} & 
 \mathbf{A}^{L_{1}/CT_{1}} & 
 \mathbf{A}^{L_{1}/CT_{2}} \\ 
 \mathbf{A}^{L_{2}/L_{1}} & 
 \mathbf{A}^{L_{2}} & 
 \mathbf{A}^{L_{2}/CT_{1}} & 
 \mathbf{A}^{L_{2}/CT_{2}} \\ 
 \mathbf{A}^{CT_{1}/L_{1}} & 
 \mathbf{A}^{CT_{1}/L_{2}} & 
 \mathbf{A}^{CT_{1}} & 
 \mathbf{A}^{CT_{1}/CT_{2}} \\
 \mathbf{A}^{CT_{2}/L_{1}} & 
 \mathbf{A}^{CT_{2}/L_{2}} & 
 \mathbf{A}^{CT_{2}/CT_{1}} & 
 \mathbf{A}^{CT_{2}} \\ 
 \end{pmatrix} 
\begin{pmatrix} 
\mathbf{X}^{L_{1}} \\
\mathbf{X}^{L_{2}} \\
\mathbf{X}^{CT_{1}} \\
\mathbf{X}^{CT_{2}} \\

\end{pmatrix} = \bm{\omega}
\begin{pmatrix} 
\mathbf{X}^{L_{1}} \\
\mathbf{X}^{L_{2}} \\
\mathbf{X}^{CT_{1}} \\
\mathbf{X}^{CT_{2}} \\
\end{pmatrix}, \label{subTDA}
\end{equation}
\end{widetext}

We can now identify couplings between local and charge-transfer excitations as off-diagonal blocks of the $\mathbf{A}$ matrix.\cite{bensberg2019automatic,tolle2019inter}
Thanks to the re-canonicalization and equation (\ref{Flocal}), the diagonal blocks have the same structure as in the CMO basis, i.e. Eq. (\ref{A_CMO}). 
For the off-diagonal blocks, we need to employ the more general expression (\ref{A}) to account for Eqs. (\ref{FCT1}) and (\ref{FCT2}). 
Without truncation the results in the RILMO basis will be identical to those obtained in the CMO basis as we will numerically verify in section \ref{CD_vs_RD}. Obviously, this case is not of practical interest, as one aims to reduce the computational cost of the method.
Hence, we consider two subsequent truncations that will each affect the results:
(i) reduction of the size of the RILMO basis by limiting the total number of virtual orbitals $N_v$ that are localized and (ii) reduction of the number of couplings between subsystems that are accounted for by restricting the solution of the TDA equation to only one or a few subsets of excitations. 
In the following, we refer to applying (ii) as the `reduced diagonalization' (RD), as opposed to the `complete diagonalization' (CD) when (ii) is not applied.
One should keep in mind that the truncation (i) can be so severe that even the CD results will significantly deviate from the full CMO treatment. 
We will therefore first study the effect of this initial truncation (i), referred to as the ``RILMO truncation''.

\subsection{Reduced diagonalization}
As the number of solutions sought is typically far smaller than the dimension of the $\textbf{A}$ matrix, it is advantageous to use a Davidson diagonalization algorithm.\cite{er1975iterativecalculationof}
We can thereby make use of the partitioning, and solve each particular subset $S$ of excitations individually, initially without accounting for couplings between the subsets,
\begin{equation}
\mathbf{A}^{S} \mathbf{V}^{S} = 
\bm{\omega}^S
\mathbf{V}^{S}.
\end{equation}

\begin{figure*}
\centering
\includegraphics[width=0.9\textwidth]{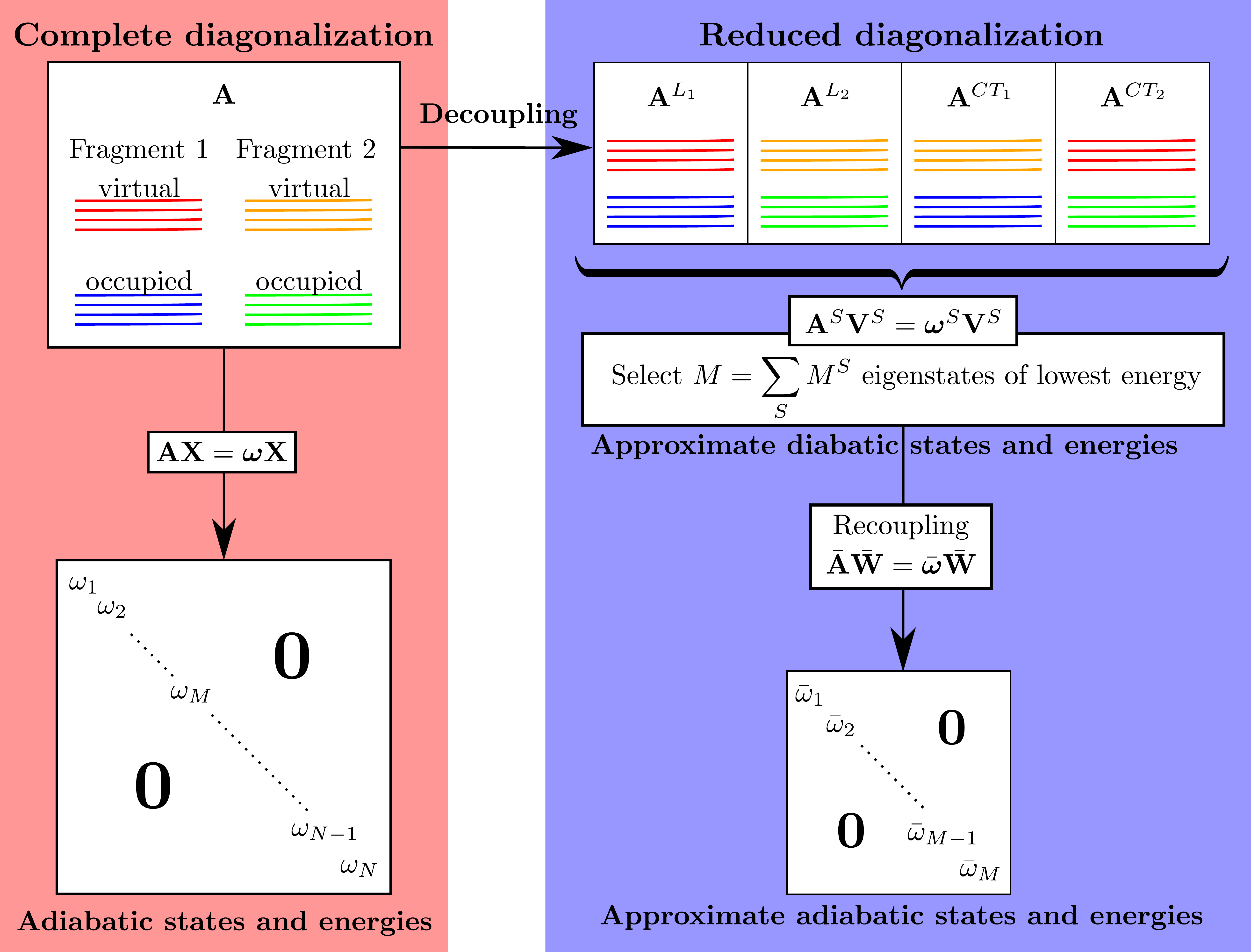}
\caption{RD procedure for obtaining the LE and CT states and electronic couplings using RILMOs.} \label{roadmap}
\end{figure*}

For a partitioning into two subsystems, the eigenvectors $ \mathbf{V}_\gamma^{S}$ associated with the excitation energies $\omega_\gamma^{S}$ can be interpreted as approximate diabatic states corresponding to $L_1$, $L_2$, $CT_1$ and $CT_2$. 
They are subsequently coupled to form adiabatic states via
\begin{equation}
\mathbf{\bar{A}} \mathbf{\bar{W}} = \bar{\bm{\omega}} \mathbf{\bar{W}},
\label{reduced_TDA}
\end{equation}
with the matrix elements of $\mathbf{\bar{A}}$ defined as
\begin{align}
\bar{A}^{S}_{\gamma\delta} &= \delta_{\gamma\delta}\omega^{S}_\gamma \\
\bar{A}^{S/T}_{\gamma\delta} &= \sum_{ia \in S}^{} \sum_{jb \in T}^{} 
V^{S}_{ia,\gamma}
A^{S/T}_{ia,jb}
V^{T}_{jb,\delta},
\label{coupling}
\end{align} 
yielding the coefficient vector $\mathbf{\bar{W}}$ which expresses the adiabatic state in terms of the diabatic states. 
This approach can be generalized to many subsystems, with the efficiency arising from the fact that only a small number $M^{S}$ of lowest eigenvectors for each excitation subset $S$ is used.  
The resulting $M=\sum_{S}{M^{S}}$ dimensional matrix problem for the full system is then easily solvable. 
The calculation of the electronic couplings between subsystems and subsequent diagonalization in a reduced space has a similar structure as encountered in the subsystem TDA approach, as discussed in Refs.\citenum{konig2013direct} and \citenum{tolle2020electronic}. 
Hence, RD consists in constructing and subsequently diagonalizing (see Eq.~\ref{reduced_TDA}) a basis of approximate diabatic states, where only a limited set of the lowest diabatic states per subspace is considered.
The RD pathway for a case of two subsystems is schematically depicted in Fig.~\ref{roadmap} and can be summarized as:
\begin{enumerate}
    \item Starting from the set of RILMOs, partition the occupied and virtual orbitals and perform four separate TDA calculations to obtain sets of eigenvectors $\{\mathbf{V}_{\gamma}^{S}\}$ for each of the $L_1$, $L_2$, $CT_1$ and $CT_2$ sub-spaces.
    \item Calculate electronic couplings between the diabatic states using the sets of eigenvectors from the previous step and construct matrix $\mathbf{\bar{A}}$, 
    \item Diagonalize $\mathbf{\bar{A}}$ to obtain the adiabatic states.
\end{enumerate} 

Note that in the limit of including all the diabatic states for each subspace, the CD and RD procedures yield identical results.

\subsection{Evaluation of the coupling elements}

The most time-consuming steps in the calculation are the matrix-vector multiplications, in which trial vectors are contracted with a block of the $\mathbf{A}$ matrix. 
We will briefly summarize our implementation, which is based on the pair-fitting approach of the ADF engine of the Amsterdam Modelling Suite (AMS), as described in Ref.~\citenum{van1999implementation}. 
We first define the RILMOs in terms of atomic orbitals $\chi$ as
\begin{eqnarray}
\phi_{l}(\mathbf{r}) = \sum_{\kappa} C^{RILMO}_{\kappa l} \chi_{\kappa}(\mathbf{r}),
\end{eqnarray}
and introduce the density functions $f_{p}$ to fit the transition densities $\rho^T_{\delta}(\mathbf{r})$ of state $\delta$ of subspace $T$
\begin{eqnarray}
\rho^T_{\delta}(\mathbf{r}) = \sum_{jb \in T} \phi_{j}(\mathbf{r})\phi_{b}(\mathbf{r})V_{jb,\delta}^{T} \approx \sum_{p}f_{p}(\mathbf{r})c^T_{p\delta} . \label{k3} 
\end{eqnarray}

The fit coefficients $c^T_{p\delta}$ can be expressed in terms of atomic orbital fit coefficients $\mathbf{c}^{AOfit}$ and the (trial) $\mathbf{V}$ matrices as
\begin{eqnarray}
c^T_{p\delta}=\sum_{jb \in T} \sum_{\kappa, \lambda} C^{RILMO}_{\kappa j}C^{RILMO}_{\lambda b}c^{AOfit}_{\kappa\lambda,p}V_{jb,\delta}^{T}.
\end{eqnarray}
Using Eq. (\ref{k3}), the induced potential $\delta v_{ind,\delta}^{T}$ for state $\delta$ of subspace $T$ can then be written as
\begin{eqnarray}
&\delta v_{ind,\delta}^{T}(\mathbf{r}) &= \sum_{p} \big [ g_{p}(\mathbf{r}) + f_{xc}(\mathbf{r}) f_{p}(\mathbf{r}) \big ]c^T_{p\delta},
\label{fit1}
\end{eqnarray}
with $g_{p}(\mathbf{r}) $ the Coulomb potential of the fit function $f_{p}(\mathbf{r})$ and $f_{xc}(\mathbf{r})$ the exchange-correlation kernel. 
The matrix representation of the induced potential in the RILMO excitation basis of subspace $S$ can subsequently be obtained by numerical integration
\begin{eqnarray}
&[\delta v^{T}_{ind,\delta}]_{ia}^{S} &= \sum_{k} w_{k} \phi_{i}(\mathbf{r}_k)\delta v_{ind,\delta}^{T}(\mathbf{r}_k) \phi_{a}(\mathbf{r}_k),
\label{fit2}
\end{eqnarray}
with $w_{k}$ the weight associated to the grid point $\mathbf{r}_k$. 
Because the functions $\phi_i$ and $\phi_a$ are local, detection and neglect of small contributions to the integrand can be used to speed up this step.

For hybrid functionals, we also need to consider the exchange contribution. 
Here, density fitting can be used as well, using the implementations reported in Refs. \citenum{franchini2014accurate} and \citenum{forster2020quadratic}. 
The exchange contribution to the induced potential is given by 
\begin{equation}
[\delta v^T_{exc,\delta}]_{ia}^S=\sum_{jb \in T} (ij|ab) V^T_{jb,\delta},
\end{equation}
which is first expressed in the AO basis and evaluated using density fitting as
\begin{align}
[\delta v^T_{exc,\delta}]_{\kappa\lambda} &= \sum_{\mu\nu} (\kappa \mu | \lambda \nu)  V^T_{\mu\nu,\delta} \nonumber
\\
 &\approx \sum_{\mu\nu} \sum_{p,q} c^{AOfit}_{\kappa\mu,p} c^{AOfit}_{\lambda\nu,q} (f_{p}|f_{q})V^T_{\mu\nu,\delta} ,\label{exchange},
\end{align}
with
\begin{equation}
V^T_{\mu\nu,\delta}=\sum_{jb \in T} C^{RILMO}_{\mu j}C^{RILMO}_{\nu b}
V^T_{jb,\delta}. 
\end{equation}
After a transformation to the excitation subspace $S$
\begin{equation}
[\delta v^T_{exc,\delta}]_{ia}^{S} =\sum_{\kappa\lambda} 
C^{RILMO}_{\kappa i}C^{RILMO}_{\lambda a}
[\delta v^T_{exc,\delta}]_{\kappa\lambda},
\end{equation}
this term can be combined with the other contributions. 
The total
electronic coupling in Eq. (\ref{coupling}) between two excitations $\gamma$ and $\delta$ belonging to the $S$ and $T$ subspaces then becomes
\begin{align}
\bar{A}^{S/T}_{\gamma\delta} &=\sum_{ia \in S} \sum_{jb \in T} V_{ia,\gamma}^{S} (F^{RILMO}_{ij}\delta_{ab}-F^{RILMO}_{ab}\delta_{ij})V^T_{jb,\delta} \nonumber \\
&+ \sum_{ia \in S} V_{ia,\gamma}^{S} \bigg ( [\delta v^{T}_{ind,\delta}]_{ia}^{S} - c_{x}[\delta v^T_{exc,\delta}]_{ia}^{S}
\bigg ). \label{k4}
\end{align}

\subsection{Tight-binding approximations to the coupling matrix}

The above-mentioned scheme can also be easily applied to approximate TDDFT approaches. 
An example is Grimme's simplified TDA\cite{grimme2013simplified} in which matrix elements of the coupling matrix $\mathbf{K}$ are given as
\begin{equation}
K_{ia,jb} = \sum_{A,B}^{N_{atoms}}(2q_{ia}^{A}\gamma_{AB}^{J}q_{jb}^{B} - q_{ij}^{A}\gamma_{AB}^{K}q_{ab}^{B}), \label{sTDA}
\end{equation}
with atomic transition charges $q_{pq}^{A}$ obtained as
\begin{eqnarray}
q_{pq}^{A} = \sum_{\mu \in A}C'_{\mu p}C'_{\mu q}.
\label{transition_charges}
\end{eqnarray}
The parameters $\gamma_{AB}^{J}$ and $\gamma_{AB}^{K}$ depend on the interatomic distance and the chemical hardness of the atoms as well as on a few empirical parameters. 
The matrix $\mathbf{C}'$ denotes MO coefficients in the orthogonal AO basis obtained from 
$\mathbf{C'}  = \mathbf{S}^{1/2}\mathbf{C}$ with $\mathbf{C}$ the coefficients in the original AO basis and $\mathbf{S}$ the AO overlap matrix. 
This approach is easily adapted to the RILMO partitioning by just replacing the CMOs by the RILMOs in Eqs. (\ref{sTDA}) and (\ref{transition_charges}).

\section{Computational Details}

All DFT and TDDFT calculations using the RILMOs were carried out using a locally modified version of the Amsterdam Modelling Suite of Programs \cite{te2001chemistry,AMS2021} (AMS). 
For all TDDFT calculations in AMS, the ALDA kernel and TDA were used throughout this work (denoted as RILMO-TDA).
The localization and re-canonicalization of the molecular orbitals were carried out  separately using a recently developed stand-alone program, so-called Reduction of Orbital Space Extent (ROSE) that has an interface to AMS\cite{senjean2021generalization,rose}.
All calculations for ROSE and AMS were automated using the Python Library for Automating Molecular Simulation \cite{PLAMS2021} (PLAMS) of AMS. 
The sorting of the orbitals before the re-canonicalization step was done based on a Mulliken population analysis of the LMOs over each fragment.
In this work we used the same level of basis sets, $\mathcal{B}_{1}$ and $\mathcal{B}^{k}$, for the supermolecule and the $k$ fragments respectively (see reference \citenum{senjean2021generalization} for the notations). 
For the localization step, we used the Pipek--Mezey localization for both the occupied and the virtual orbitals.\cite{pipek1989fast}
For TDDFT calculations with the Grimme's simplified Tamm--Dancoff \cite{grimme2013simplified} approximation (denoted as RILMO-simTDA),  the parameters from Ref.~\citenum{grimme2013simplified},  corresponding to the standard settings in AMS, were used. \\
All projection-based embedding subsystem DFT (PbE-sDFT) and TDDFT/TDA calculations (denoted as PbE-sTDA) calculations were carried out using SERENITY version 1.4.0 \cite{serenity_pub} with the LevelShift  projection \cite{manby2012simple} operator of the top-down embedding \cite{tolle2019inter,bensberg2019automatic,scholz2020analysis} procedure as outlined in Ref.~ \citenum{bensberg2019automatic}. 
Additional localization of the virtual orbitals was carried out using the approach outlined in Ref.~\citenum{scholz2020analysis} 
as implemented inside SERENITY. 
In all the calculations performed in SERENITY, the resolution of identity (RI) \cite{eichkorn1995auxiliary} approximation was used for the evaluation of the Coulomb contribution in all the DFT and TDDFT/TDA \cite{weigend2002efficient} calculations.

\section{Results and Discussions}

\subsection{RILMO truncation and Complete versus Reduced Diagonalization}
\label{CD_vs_RD}

\begin{figure}
\centering
\includegraphics[width=0.5\textwidth]{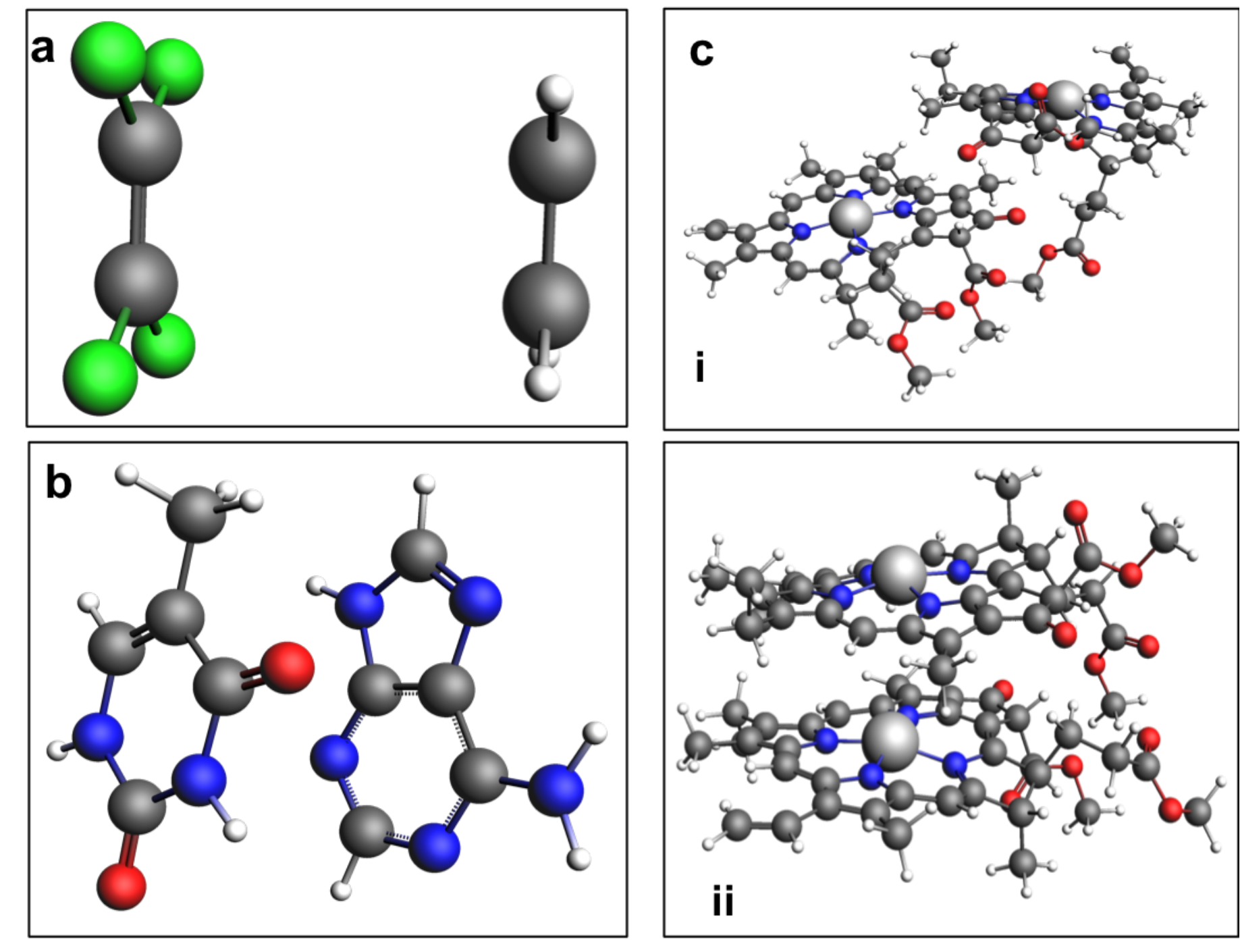}
\caption{Geometries of the a) tetrafluoroethylene-ethylene dimer, b) adenine-thymine base pair, and c) chlorophyll dimer.}\label{Structures}
\end{figure}

In order to demonstrate our procedure, we choose two test systems: the ethylene-tetrafluoroethylene dimer and the adenine-thymine DNA base pair. 
While the former presents an adequate model for studying LE and CT type transitions, the latter provides a biologically relevant model system where such quantities are of  importance to model the stability of biomolecules under irradiation\cite{blancafort2014exciton,improta2016quantum}.
The geometry of both of these structures were taken from Ref.~\citenum{tolle2020electronic} and Ref.~\citenum{blancafort2014exciton} and is displayed in Fig.~\ref{Structures}a and \ref{Structures}b, respectively. The CAM-B3LYP\cite{yanai2004new} exchange-correlation functional along with the Slater-type basis set TZ2P as employed in AMS was used for both these systems. 
 \\

\begin{figure*}
\centering
\includegraphics[width=0.9\textwidth]{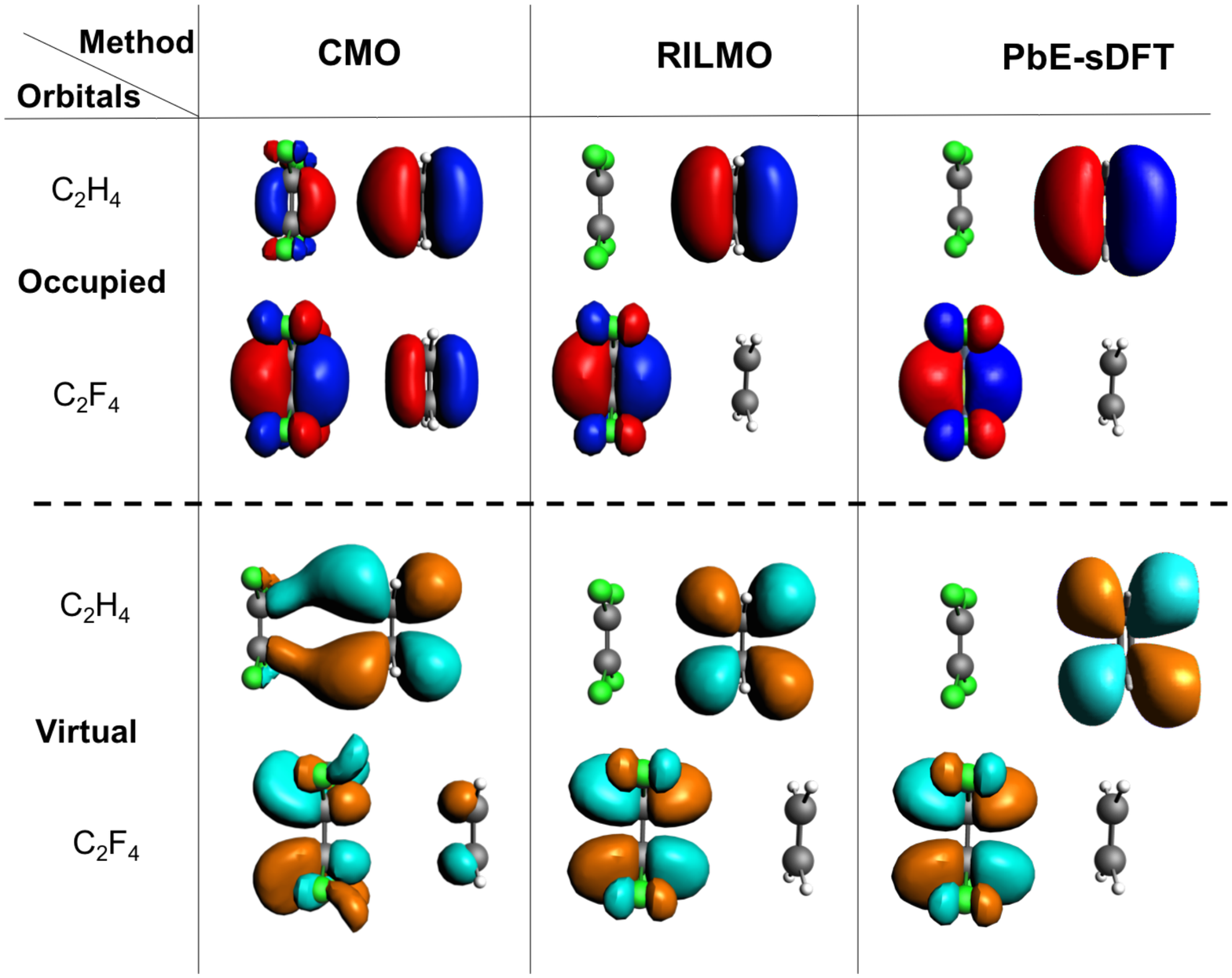} \\
\caption{The occupied and virtual orbitals associated with the uLE and uCT states for the $C_{2}H_{4}-C_{2}F_{4}$ dimer obtained from CMOs (left column), RILMOs (middle column) and PbE-sDFT (right column).}\label{MOs_C2h4C2f4}
\includegraphics[width=0.9\textwidth]{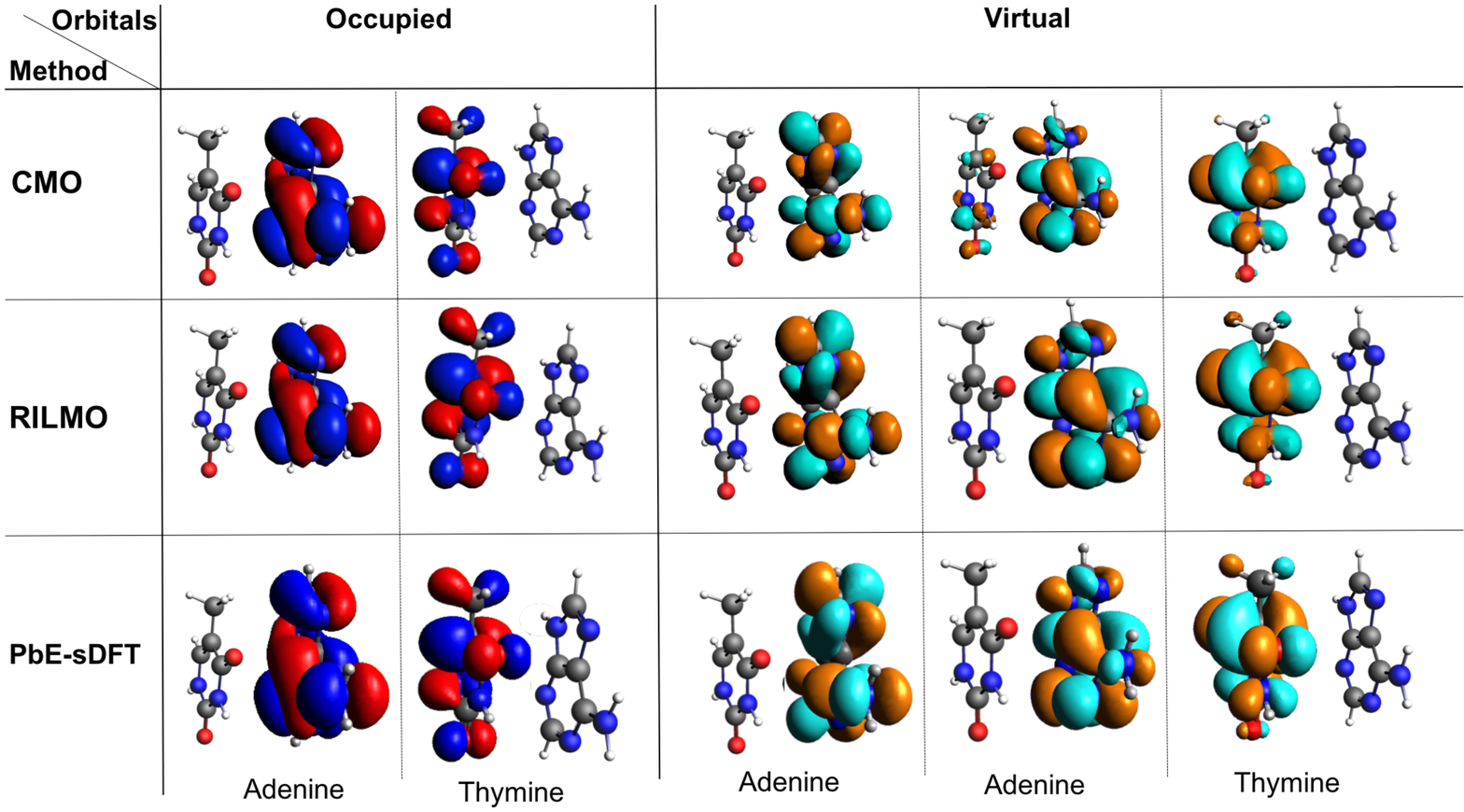}
\caption{The occupied and virtual orbitals associated with the uLE and uCT states for the adenine-thymine dimer obtained from CMOs (top row), RILMOs (middle row) and PbE-sDFT (bottom row).} \label{MOs_AT}
\end{figure*}

The uncoupled LE (uLE) and uncoupled CT (uCT) states considered for both systems consist of $\pi \rightarrow \pi^{*}$ type transitions. 
Fig.~\ref{MOs_C2h4C2f4} (middle column) and Fig.~\ref{MOs_AT}  (middle row) plot the relevant $\pi$ and $\pi^{*}$ RILMOs associated with the $C_{2}H_{4}-C_{2}F_{4}$ dimer and the adenine-thymine dimer, respectively. 
The corresponding supermolecular CMOs are also shown. 
The CMOs are delocalized over both fragments, and the localization and re-canonicalization procedures restore the typical $\pi$ and $\pi^{*}$ character in the RILMOs. For comparison, we show the PbE-sDFT orbitals that yield a similar localized picture.
\begin{figure*}
\centering
\includegraphics[width=0.9\textwidth]{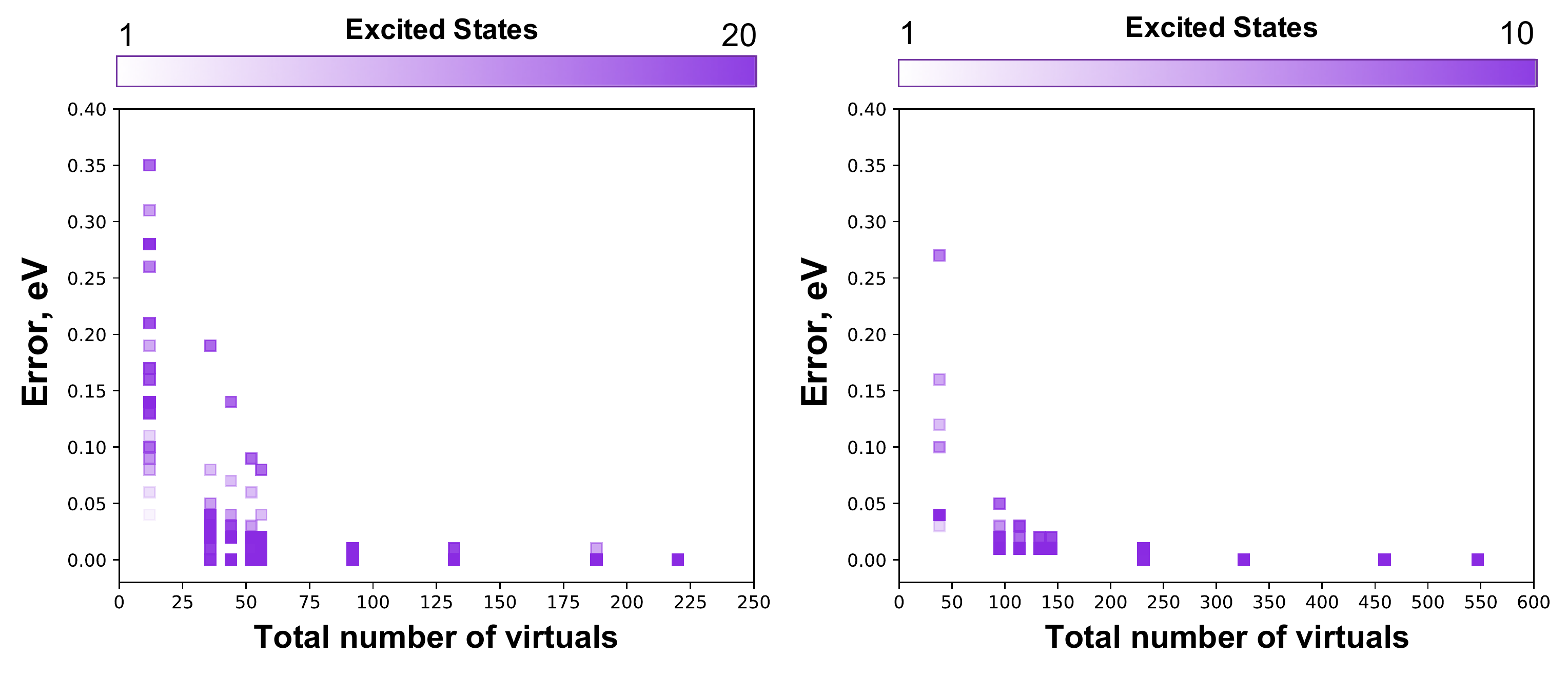}
\caption{Error in excitation energy (as compared to the supermolecular CMO reference energies) due to RILMO truncation of hv orbitals. Left panels: lowest 20 RILMO-TDA (CD) states for the $C_{2}H_{4}-C_{2}F_{4}$ dimer. Right panels: lowest 10 RILMO-TDA (CD) states of the adenine-thymine dimer. $M$ denotes the number of diabatic states (top panels: $M=20$, bottom panels: $M=80$). The color of the marker indicates the energy ordering of states.}\label{E_CMO_RILMO_CD_1}
\end{figure*}
\begin{figure*}
\centering
\includegraphics[width=0.9\textwidth]{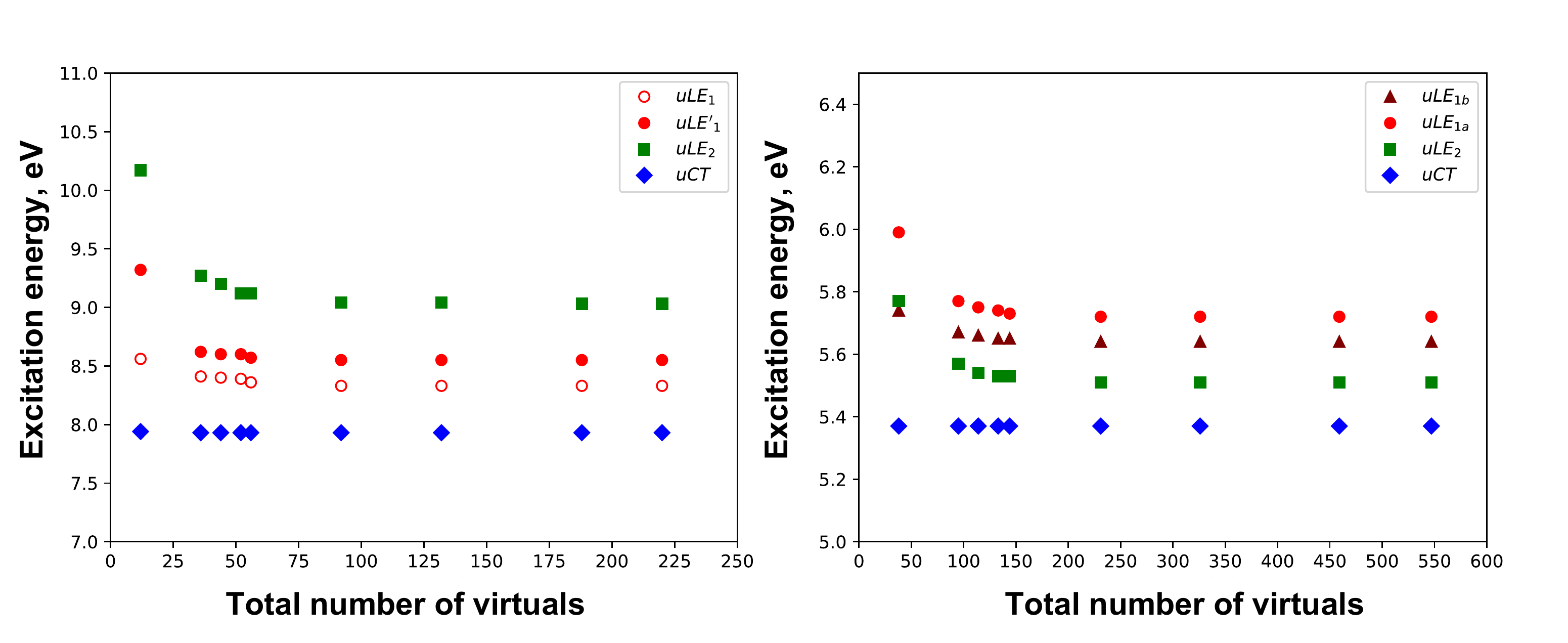} \\
\caption{Left panel: Excitation energy of the uLE and uCT states using RILMO-TDA (RD) for the $C_{2}H_{4}-C_{2}F_{4}$ dimer
with respect to the number of hv orbitals used in the localization procedure. Right panel: same as the left panel for the adenine-thymine dimer.}\label{E_diabatic_1}
\end{figure*}

For the ethylene-tetrafluoroethylene dimer we consider two uLE states for $C_{2}H_{4}$, one uLE for $C_{2}F_{4}$ and one uCT state from $C_{2}F_{4} \rightarrow C_{2}H_{4}$. For $C_{2}H_{4}$ we have a $\pi \rightarrow \pi^{*}$ state and a $\pi \rightarrow \sigma^{*}$ state which is near-degenerate with the former. These two states mix upon increasing the size of the virtual space in the localization procedure, so that we will denote them as uLE$_{1}$ and uLE$_{1}^{'}$ below.
In the limit of the full virtual space, the uLE$_{1}$ retains mostly the $\pi \rightarrow \pi^{*}$ character ($\sim$ 71 \%). For $C_{2}F_{4}$ we have a $\pi \rightarrow \pi^{*}$ state denoted as uLE$_{2}$ and for the CT transition we have a $ C_{2}F_{4}, \pi \rightarrow C_{2}H_{4}, \pi^{*}$ state denoted as uCT. 
For the adenine-thymine dimer, we label the two near-degenerate adenine $\pi \rightarrow \pi^{*}$ states (originally denoted L$_{a}$ and L$_{b}$ in Ref.~\citenum{blancafort2014exciton}) as uLE$_{1a}$ and uLE$_{1b}$, while denoting the thymine $\pi \rightarrow \pi^{*}$ state as uLE$_{2}$. 
Besides these three uLE states, there is one uCT state: $ \textrm{adenine}, \pi \rightarrow \textrm{thymine},\pi^{*}$. It is important to note that the above set of four excitations for both of these dimers does not constitute the lowest set of excitations but selects the primary valence $\pi \rightarrow \pi^{*}$ type transitions that are often used to model LE and CT type excitations in the literature.

\begin{figure*}
\centering
\includegraphics[width=0.9\textwidth]{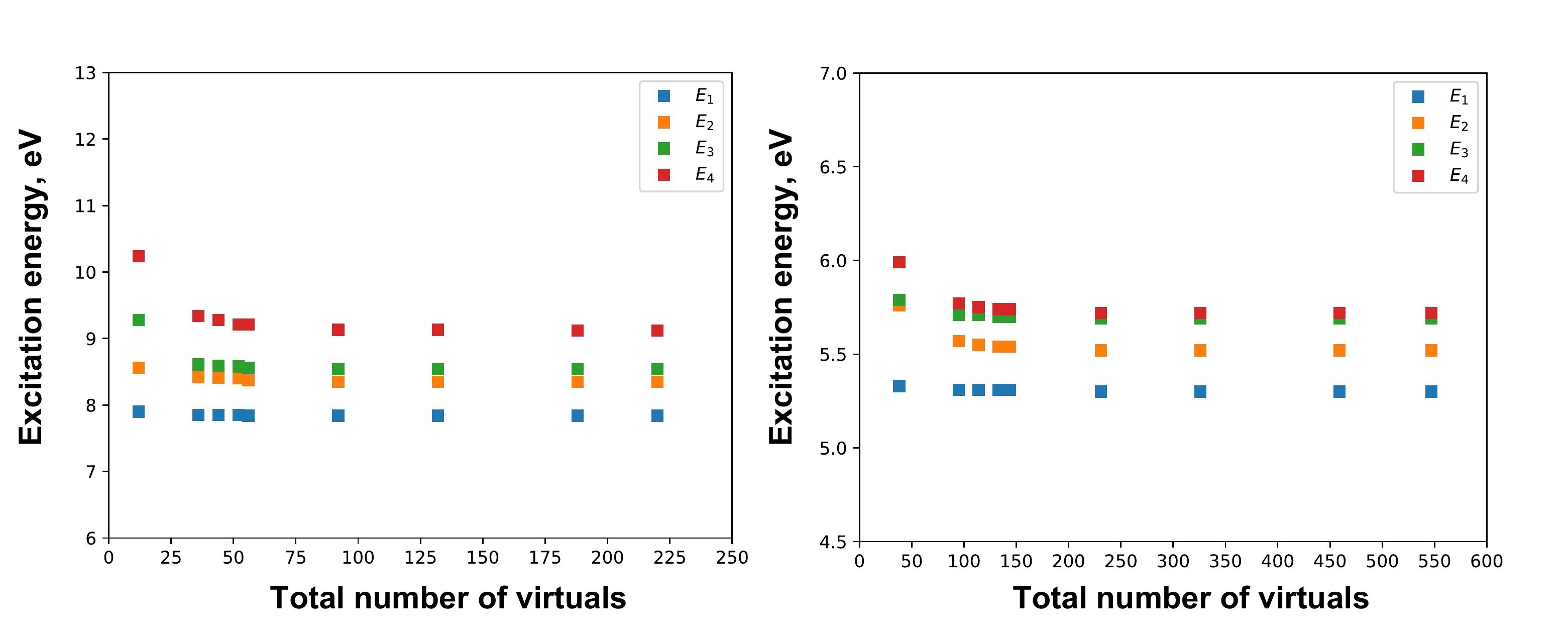} \\
\caption{Excitation energies
of the coupled states using the minimal set of four diabatic states for the $C_{2}H_{4}-C_{2}F_{4}$ (left panel) and adenine-thymine (right panel) dimers
with respect to the number of hv orbitals used in the localization procedure. }\label{E_RILMO_RD_4}
\end{figure*}

\begin{figure*}
\centering
\includegraphics[width=0.9\textwidth]{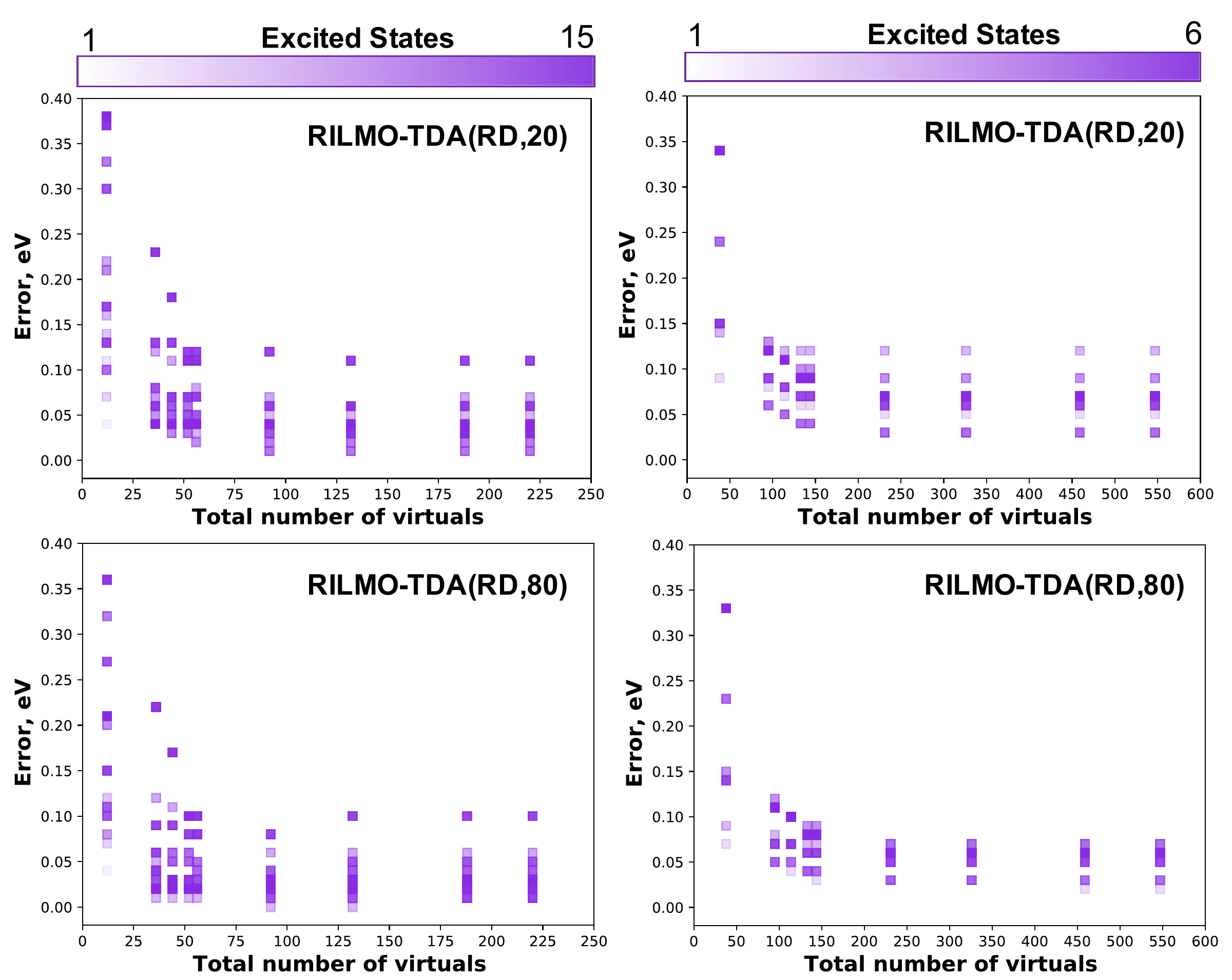} \\
\caption{Error in excitation energy (as compared to the supermolecular CMO reference energies) due to RILMO truncation of hv orbitals. Left panels: lowest 15 RILMO-TDA (RD,$M$) states for the $C_{2}H_{4}-C_{2}F_{4}$ dimer. Right panels: lowest 6 RILMO-TDA (RD,$M$) states of the adenine-thymine dimer. $M$ denotes the number of diabatic states (top panels: $M=20$, bottom panels: $M=80$). The color of the marker indicates the energy ordering of states.}\label{E_RILMO_RD_x}
\end{figure*}

As a numerical verification of our implementation, we first show that the supermolecular excitation spectrum arising from the CD (see Fig.~\ref{roadmap}) of RILMO-TDA converges to the CMO results in the limit of using the full virtual space.
Fig.~\ref{E_CMO_RILMO_CD_1} plots the error in the energy of the lowest RILMO-TDA states from the CMOs as a function of the virtual space size, i.e from severe to no RILMO truncation. Besides the expected convergence, the plot also indicates that RILMO truncation is a viable way to restrict the excitation space. A modest amount of virtual orbitals suffices to obtain adiabatic state energies close to the reference supermolecular CMO results. 

Next, we consider combining RILMO truncation with truncation in the number of diabatic states employed in the RD scheme. For this purpose, we couple for both dimers only the aforementioned sets of four uLE and uCT states.
Fig.~\ref{E_diabatic_1} shows the effect of RILMO truncation combined with RD. From Fig.~\ref{E_diabatic_1}, we observe a decrease in the uLE energies for both the $C_{2}H_{4}-C_{2}F_{4}$ and adenine-thymine dimers, whereas the uCT energies are quite stable upon increasing the number of virtual orbitals. 
This stability can be explained by the fact that the CT excitations are well described by single orbital transitions,\cite{sen2022towards} so that moving from a minimal valence space to the full virtual space causes little to no change in the corresponding excitation energies.
The LE excitations, on the other hand, are mixtures of local single orbital transitions and are thus more sensitive to the number of virtual orbitals.
The uLE excitation energies exhibit a smooth convergence upon increasing the number of virtual orbitals, similar to what is seen for isolated or well-separated fragments (see SI, Figs.~S1 and S2). 

Fig.~\ref{E_RILMO_RD_4} plots the corresponding coupled adiabatic states using the above four diabatic states for both of these dimers. As can be seen, the curves follow closely the uncoupled energies and converge smoothly upon increasing the total number of virtual orbitals. Figs.~\ref{E_diabatic_1} and \ref{E_RILMO_RD_4} show that it is sufficient to work with a modest number of virtual orbitals to construct the local states in the RD framework.

Having thus shown that RILMO truncation in combination with RD is viable, we also consider the RD procedure itself in more detail. To do so, we couple the above minimal set of four diabatic states with additional low-lying states and investigate how this changes the excitation energies. Fig.~\ref{E_RILMO_RD_x} plots the error in the excitation energies of the lowest RILMO-TDA adiabatic states from their respective CMO states. We thereby chose a simple set up, taking either the lowest 5 or the lowest 20 diabatic states in each of the four excitation subspaces, yielding a total of 20 and 80 adiabatic states, respectively. 
A direct comparison of all resulting states with CMO states is complicated, as the coupling of adiabatic states also gives rise to high-energy states that can not be readily matched with CMO states. 
For this reason, we focus on the lowest energy states up to the above-mentioned bright $\pi \rightarrow \pi^{*}$ excitations. 
As can be seen for both of these dimers, increasing the total number of states from 20 to 80 shows a reduction in the overall deviation of the lower-lying states to well within one-tenth of an eV, again already with a modest amount of virtual orbitals (i.e. with strong RILMO truncation). 
From Fig.~\ref{E_RILMO_RD_x}, we conclude that for the $C_{2}H_{4}-C_{2}F_{4}$ and adenine-thymine dimers, a set of five states for each of the subspaces is sufficient to closely reproduce the lowest CMO states, including the characteristic $\pi \rightarrow \pi^{*}$ transitions. As an additional proof of principle, we have further checked the smooth convergence of the lowest 10 RILMO-TDA adiabatic states towards the reference supermolecular CMO states, with respect to the $M$ number of diabatic states used in the RD pathway on a smaller water-ammonia test system (see Fig.~S4 in SI).

\subsection{Comparison with projection-based embedding subsystem DFT}

\begin{figure*}
\centering
\includegraphics[width=0.9\textwidth]{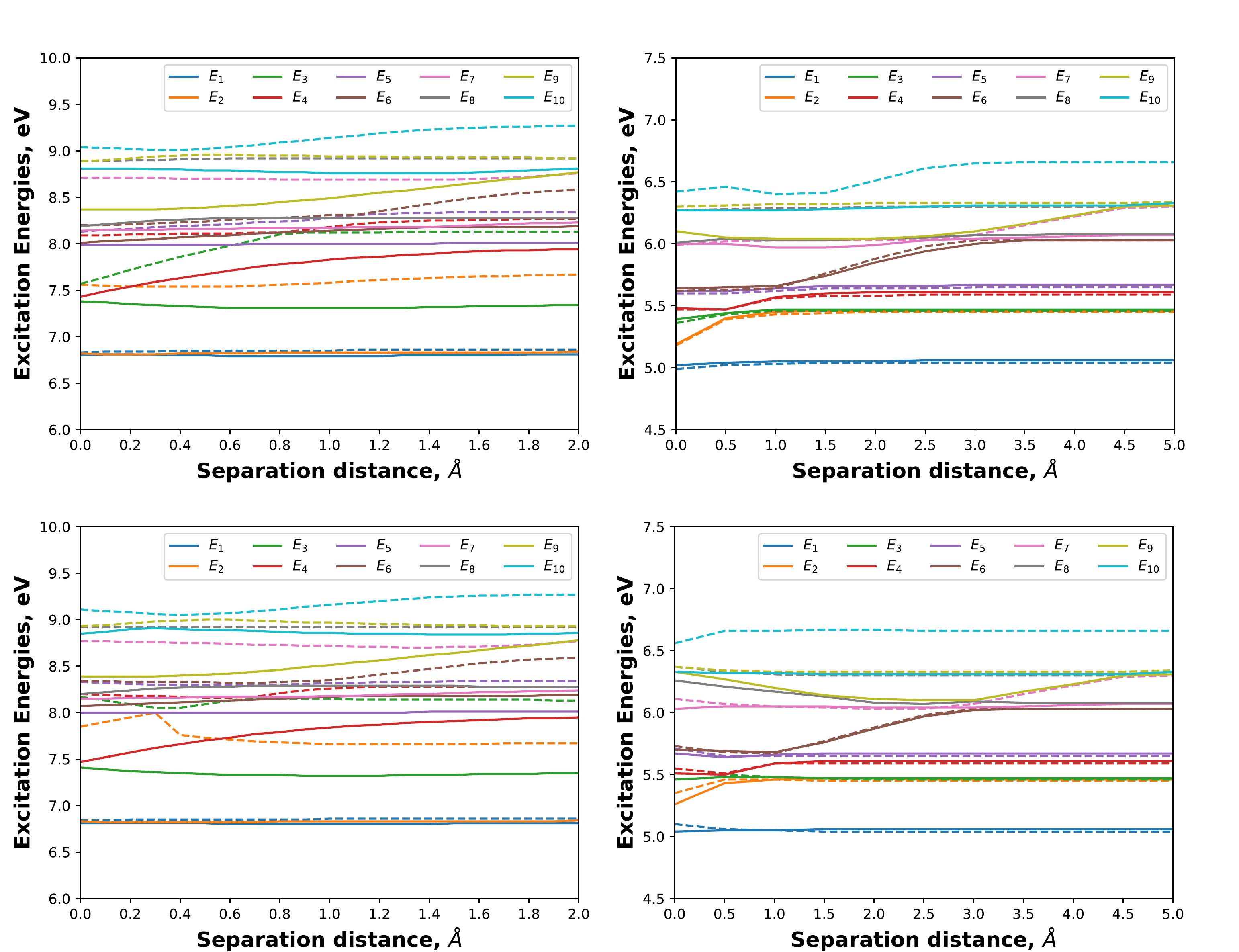}
\caption{Top panel: Excitation energies of the lowest 10 supermolecular states in the CMO basis from AMS (full lines) and SERENITY (dashed lines) using TDA with respect to the separation distance of the $C_{2}H_{4}-C_{2}F_{4}$ (left panel) and adenine-thymine (right panel) dimers. Bottom panel: Excitation energies of the lowest 10 adiabatic states obtained by RD pathway with $M= 80$ diabatic states from RILMO-TDA (full lines) and PbE-sTDA (dashed lines) for both dimers.}\label{E_CMO_AMS_SER_C2h4C2f4}
\end{figure*}

Having shown the convergence towards the supermolecular CMO results for the CD and RD pathways, we now focus on a more detailed comparison of the above approximate diabatic uLE and uCT states as well as their electronic couplings with those obtained from PbE-sDFT/sTDA. 
For the sake of consistency, we retain the full virtual space in the RILMO-TDA calculation of the above dimers. 
Both of these systems have been studied in Ref.~\citenum{tolle2020electronic} and their diabatic LE and CT states were obtained using the multistate-FCD-FED approach and PbE-sTDA by employing different orbital partition schemes for varying displacements \textbf{d} between the corresponding monomers.
In Ref.~\citenum{tolle2020electronic}, an overall decent agreement was obtained between the two methods in spite of their differences in the construction of the diabatic states. 
We refer the reader to Ref. \citenum{tolle2020electronic} for more details, and only the necessary PbE-sDFT/sTDA calculations are repeated here for comparison. 
For the purpose of this work, the Gaussian type def2-TZVP basis set was chosen in SERENITY in order to compare with the Slater-type basis set TZ2P as employed in AMS. \\
While the two basis sets are reasonably large, the effect of basis set truncation is still nonnegligible as can be seen when comparing the supermolecular CMO results of the two codes. In Fig.~\ref{E_CMO_AMS_SER_C2h4C2f4} (top panel) we plot the differences in the lowest 10 excitation energies and see that these can amount to a few tenths of eV, with the most pronounced differences occurring for the $C_{2}H_{4}-C_{2}F_{4}$ dimer. While for the Adenine-Thymine dimer our states of interest (the characteristic $\pi \rightarrow \pi^{*}$ transitions)  lies within the lowest 6 supramolecule CMO states (states 2,3,5 and 6), such an identification for the $C_{2}H_{4}-C_{2}F_{4}$ dimer becomes difficult due to the delocalized nature of the CMOs as shown in Fig.~\ref{MOs_C2h4C2f4}. 

To compare the ground-state embedded orbitals, we again refer to Fig.~\ref{MOs_C2h4C2f4} (right column) and Fig.~\ref{MOs_AT} (bottom row) for the relevant $\pi$ and $\pi^{*}$ orbitals of $C_{2}H_{4}-C_{2}F_{4}$ and the adenine-thymine dimer. We see that despite differences in the formalism, and type of basis functions employed, both embedding procedures lead to very similar local orbitals for both systems.
\begin{table*}
  \caption{uLE energies  of the $C_{2}H_{4} - C_{2}F_{4}$ dimer obtained from RILMO-TDA and PbE-sTDA methods at varying distances $\mathbf{d}$ (\AA). Also shown are the isolated fragment energies. All units are in eV.}
 \label{uLE_C2h4c2f4_1}
 \begin{ruledtabular}
  \begin{tabular}{ccccccccccc}
       & & &\vline & $\mathbf{d=0}$ & $\mathbf{d=0.5}$  &  $\mathbf{d=1.0}$ & $\mathbf{d=1.5}$ & $\mathbf{d=2.0}$  & $\mathbf{isolated}$ &\\ 
       \hline \\[-0.8em]
      &  &$\mathbf{uLE_{1}}$&\vline & 8.33 & 8.31 & 8.30 & 8.29  &  8.28 & 8.28 \\
      &\textbf{RILMO-TDA}&  & \vline & \\
      &  & $\mathbf{uLE_{2}}$& \vline & 9.03 & 9.02   & 9.00  & 9.00  & 8.99 & 8.99 &\\   
     
     \hline \\[-0.8em]
     
      &  & $\mathbf{uLE_{1}}$&\vline  & 8.49 & 8.39  & 8.34  & 8.31  & 8.29 & 8.29 &\\
      &\textbf{PbE-sTDA}&  &  \vline & \\
      &  & $\mathbf{uLE_{2}}$& \vline &  9.03 &  8.97   &  8.94  & 8.92  & 8.91 & 8.91 &\\   
     
  \end{tabular}
  \end{ruledtabular}
  \caption{uLE energies  of the adenine-thymine dimer obtained from RILMO-TDA and PbE-sTDA methods at varying distances $\mathbf{d}$ (\AA). Also shown are the isolated fragment energies. All units are in eV.}
 \label{uLE_AT_1}
 \begin{ruledtabular}
  \begin{tabular}{ccccccccccc}
       & & &\vline & $\mathbf{d=0}$ & $\mathbf{d=1.0}$  &  $\mathbf{d=2.0}$ & $\mathbf{d=3.0}$ & $\mathbf{d=4.0}$  & $\mathbf{isolated}$ &\\ 
       \hline \\[-0.8em]
      &  &$\mathbf{uLE_{1a}}$&\vline & 5.72 & 5.69 & 5.68 & 5.67  &  5.67 & 5.67 &\\
      &\textbf{RILMO-TDA}  & $\mathbf{uLE_{1b}}$& \vline & 5.64 & 5.62  & 5.61  & 5.61  & 5.61 & 5.61 &\\ 
      &  & $\mathbf{uLE_{2}}$& \vline & 5.51 & 5.48   & 5.47  & 5.47  & 5.47 & 5.48 &\\   
     
     \hline \\[-0.8em]
     
      &  &$\mathbf{uLE_{1a}}$&\vline & 5.75 & 5.67 & 5.65 & 5.65  &  5.65 & 5.65 &\\
      & \textbf{PbE-sTDA} & $\mathbf{uLE_{1b}}$& \vline & 5.69 & 5.60  & 5.59  & 5.59  & 5.59 & 5.59 &\\ 
      &  & $\mathbf{uLE_{2}}$& \vline & 5.55 & 5.47   & 5.45  & 5.45  & 5.45 & 5.46\\   
  \end{tabular}
  \end{ruledtabular}
\end{table*}

\begin{figure*}
\centering
\includegraphics[width=0.9\textwidth]{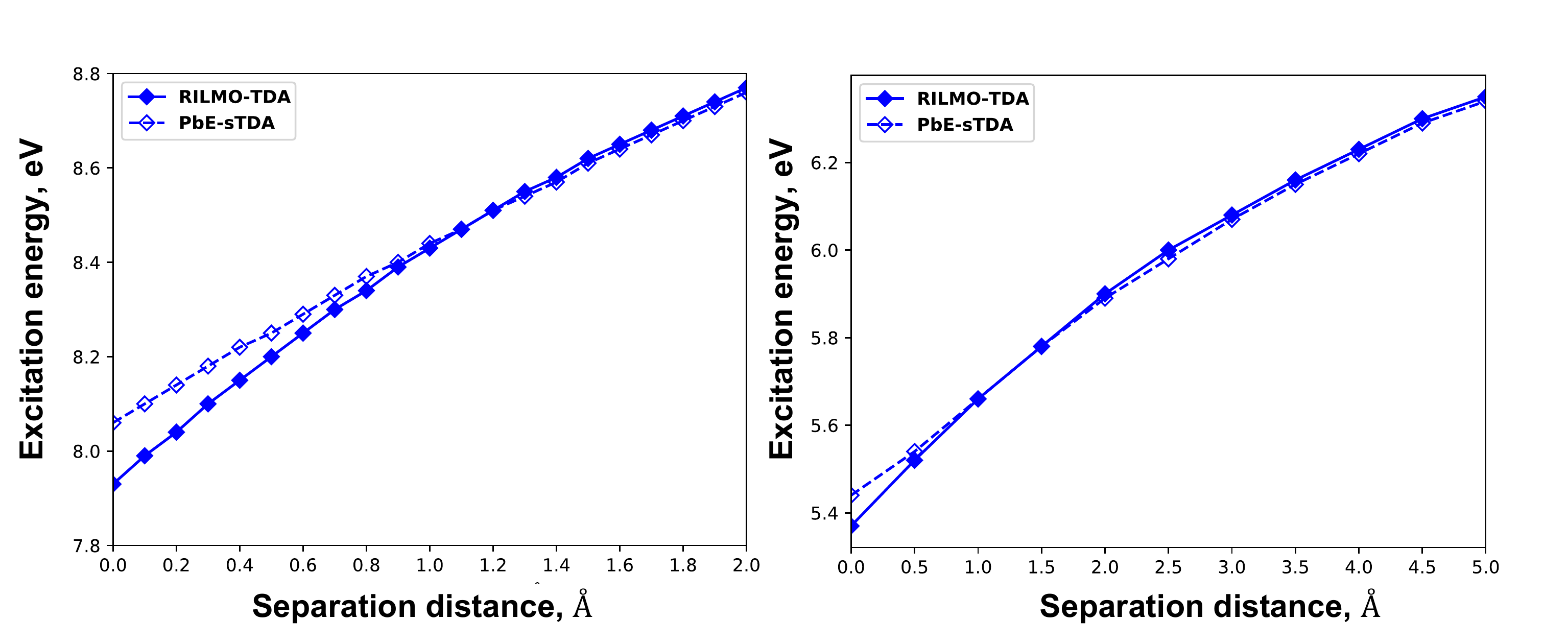}
\caption{uCT energies obtained from RILMO-TDA (full lines) and PbE-sTDA (dashed lines) as a function of the separation distance for the $C_{2}H_{4}-C_{2}F_{4}$ (left panel) and adenine-thymine (right panel) dimers.}\label{CT_c2h4c2f4andAT_1}
\includegraphics[width=0.9\textwidth]{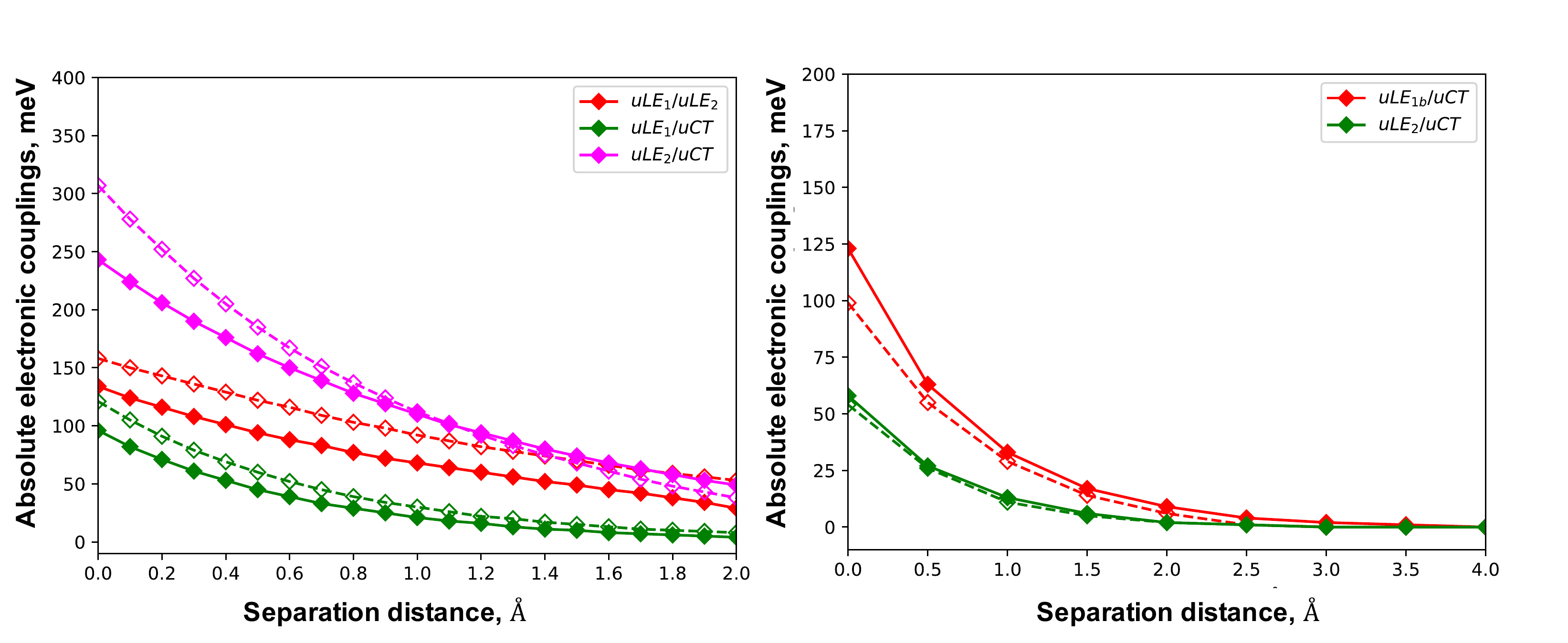}
\caption{Absolute electronic couplings between the uLE and uCT states from RILMO-TDA (full lines) and PbE-sTDA (dashed lines) as a function of the separation distance for the $C_{2}H_{4}-C_{2}F_{4}$ (left panel) and adenine-thymine (right panel) dimers.}\label{Coupling_C2H4C2H4andAT_1}
\end{figure*}

Focusing first on the lowest uCT state in each dimer, Fig.~\ref{CT_c2h4c2f4andAT_1} plots the uCT energy from RILMO-TDA along with the corresponding uCT energy from PbE-sTDA as a function of the separation distance between the monomers. 
Both methods agree very well at longer distances, displaying the asymptotic decay of $-R^{-1}$  characteristic for CT states. At shorter distances, the uLE energies from RILMO-TDA are consistently lower compared to those of PbE-sTDA (see Table~\ref{uLE_C2h4c2f4_1} for uLE$_{1}$, Table S1 in SI for the energies of uLE$_{1}^{'}$, and Table~\ref{uLE_AT_1}). 
Besides the already mentioned difference in basis sets, the deviation observed between the uCT energies could also arise from the differences in the two methods, 
in particular the different localization of the virtual orbitals.
In the PbE approach, the virtual orbitals are subject to a SVD of the MO overlap matrix computed with only the AO basis functions located on the fragment of interest to obtain virtual LMOs for that fragment\cite{scholz2020analysis,tolle2020electronic}. 
In our scheme, on the other hand, the virtual orbitals are subject to two separate PM localizations, resulting in a small set of `valence' virtual orbitals and a larger set of `hard' virtual orbitals. The resulting differences in virtual orbital energies, which in the PbE case rise more steeply upon reducing the monomer distance, appear to be the main cause for this difference (see Tables S2 and S3 in SI). 


The differences observed at short distances and agreements at long distances are also reflected in the electronic couplings between the respective uLE and uCT states. 
Fig.~\ref{Coupling_C2H4C2H4andAT_1} plots their absolute values as a function of the separation distance between the monomers for each of these dimers. 
Only a few illustrative couplings are shown, see Fig.~S5 and S6 in SI for all the couplings. 
The couplings between the uLE and uCT states calculated from both of the two methods display an exponential decay ($e^{-r}$) at long range, which is typical for exchange interactions. 
The most significant difference between the methods is seen in the uLE$_1$/uCT coupling of the $C_{2}H_{4}-C_{2}F_{4}$ dimer, where the coupling is stronger for PbE-sTDA at short distances and weaker at long distances, as compared to RILMO-TDA. 
For the other couplings, the two methods give qualitatively similar results. 

As both the PbE-sTDA and our approach converge to the supermolecular CMO results when sufficient states are coupled, we also show in Fig.~\ref{E_CMO_AMS_SER_C2h4C2f4} (bottom panel),
the excitation energies obtained when a total of 80 diabatic states are coupled via RD. 
This gives a similar picture as seen for the CMO calculations, with relatively small deviations for the adenine-thymine dimer and larger deviations for the $C_{2}H_{4}-C_{2}F_{4}$ dimer.

\subsection{Application to a \textit{chlorophyll a} dimer in Light Harvesting Complex II}

After this assessment of the RILMO-TDA approach, we now provide an application to molecules of biological importance.  
We choose a dimer consisting of chlorophyll molecules, that has been studied previously by some of us \cite{sen2021understanding} using subsystem DFT. In this work, we used the traditional version with non-additive kinetic energy functionals as opposed to projection-based embedding. 
This particular dimer is known to play a crucial role in light-induced energy transfer processes in light-harvesting complexes in higher plants \cite{sen2021understanding,liguori2015light}. 
In this previous work, we observed an increase in the mixing of charge-transfer states with the local states upon a decrease in the distance between the two chromophores. 
A direct manifestation of this mixing is the red-shifting of the lowest excited states and a decrease in the total oscillator strength. 
To study this mixing with the current method that offers an easy analysis of the effect of CT, we seek to calculate the uLE and uCT states along with their corresponding electronic couplings for two representative structures. 
In conformation 1 the distance between the two chlorophylls is relatively large, while in conformation 2 the two chlorophylls are stacked at relatively close distance (see Fig.~\ref{Structures}c). \\

The CAM-B3LYP exchange-correlation functional along with the Slater-type basis set DZP was used for this system, while we performed Gaussian-type def2-SVP with SERENITY for comparison. Here again, the full virtual space was retained (no RILMO truncation) for comparison purposes.

\begin{figure*}
\centering
\includegraphics[width=0.9\textwidth]{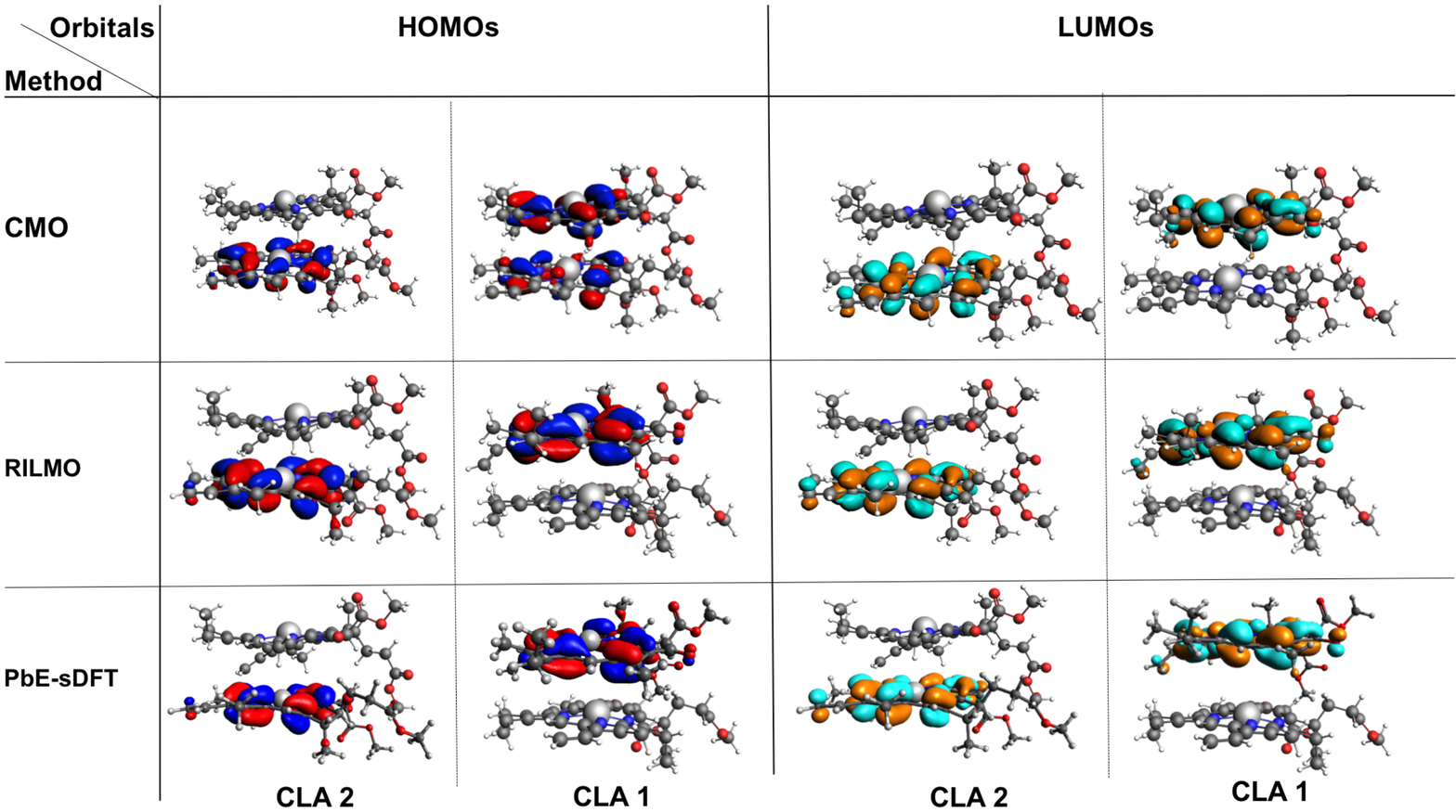}
\caption{HOMOs and LUMOs associated with the lowest $\pi \rightarrow \pi^{*}$ transition for each of the chlorophyll molecules in conformation 2(denoted as CLA1 and CLA2) from different set of orbitals: CMOs (top row), RILMOs (middle row) and orbitals from PbE-sDFT (bottom row).} \label{MOs_CLA}
\end{figure*}

\begin{table*}
  \caption{uLE and uCT energies  of the chlorophyll dimer obtained from PbE-sTDA, RILMO-TDA and RILMO-simTDA methods for both conformations. All units are in eV.}
 \label{uLE_cladimer_1}
 \begin{ruledtabular}
  \begin{tabular}{ccccccccccc}
       &  & \textbf{States} &\vline & \textbf{PbE-sTDA} & \vline &  \textbf{RILMO-TDA} & \vline & \textbf{RILMO-simTDA} &\\ 
       \hline \\[-0.8em]
      &  &$\mathbf{uLE_{1}}$ &\vline & 2.21  & \vline &2.22  & \vline & 1.93 &\\
      &  &$\mathbf{uLE_{2}}$ &\vline & 2.25  & \vline &2.26 & \vline &  1.96  &\\
      &\textbf{Conformation 1}&  & \vline & & \vline & & \vline & &\\
      &  & $\mathbf{uCT_{1}}$ &\vline & 2.82  & \vline & 2.80  & \vline & 2.36 &\\  
       &  & $\mathbf{uCT_{2}}$ &\vline & 2.97  & \vline & 2.97  & \vline & 2. 54  &\\   
     \hline \\[-0.8em]
      &  &$\mathbf{uLE_{1}}$ &\vline & 2.31  & \vline & 2.27  & \vline & 1.98 &\\
      &  &$\mathbf{uLE_{2}}$ &\vline & 2.25  & \vline & 2.25 & \vline & 1.96 &\\
      &\textbf{Conformation 2}&  & \vline & & \vline & &\vline & & \\
      &  & $\mathbf{uCT_{1}}$ &\vline & 2.70  & \vline & 2.67  & \vline & 2.22 &\\  
       &  & $\mathbf{uCT_{2}}$ &\vline & 2.30  & \vline & 2.24  & \vline & 1.79  &\\  
  \end{tabular}
  \end{ruledtabular}
  \caption{Absolute electronic couplings between the uLE/uCT states  of the chlorophyll dimer obtained from PbE-sTDA, RILMO-TDA and RILMO-simTDA methods for both conformations. All units are in meV.}
  \label{couplings_cladimer_1}
  \begin{ruledtabular}
  \begin{tabular}{ccccccccccc}
       &  & States &\vline & \textbf{PbE-sTDA} & \vline &  \textbf{RILMO-TDA} & \vline & \textbf{RILMO-simTDA} &\\ 
       \hline \\[-0.8em]
      &  &$\mathbf{uLE_{1}/uLE_{2}}$ &\vline & 27 & \vline & 26 & \vline & 35 &\\
      &  &$\mathbf{uLE_{1}/uCT_{1}}$ &\vline & 7 & \vline & 8 & \vline & 9 &\\
      &\textbf{Conformation 1}  &$\mathbf{uLE_{1}/uCT_{2}}$ &\vline & 20 & \vline & 20 & \vline & 20 &\\
      &  & $\mathbf{uLE_{2}/uCT_{1}}$ &\vline & 19 & \vline & 17 & \vline & 17 &\\  
       &  & $\mathbf{uLE_{2}/uCT_{2}}$ &\vline & 7  & \vline & 8  & \vline &9 &\\   
     \hline \\[-0.8em]
     
       &  &$\mathbf{uLE_{1}/uLE_{2}}$ &\vline & 36  & \vline & 36 & \vline &44 &\\
      &  &$\mathbf{uLE_{1}/uCT_{1}}$ &\vline & 27 & \vline & 29 & \vline & 28 &\\
      &\textbf{Conformation 2}  &$\mathbf{uLE_{1}/uCT_{2}}$ &\vline & 44 & \vline & 44 & \vline & 45 &\\
      &  & $\mathbf{uLE_{2}/uCT_{1}}$ &\vline & 31 & \vline & 29 & \vline & 37 &\\  
       &  & $\mathbf{uLE_{2}/uCT_{2}}$ &\vline & 23 & \vline & 21 & \vline & 19 &\\   
  \end{tabular}
  \end{ruledtabular}  
\end{table*}

The RILMOs and orbitals from PbE-sDFT corresponding to the HOMO ($\pi$) and LUMO ($\pi^{*}$) for the two chlorophyll molecules in conformation 2 are shown in Fig.~\ref{MOs_CLA}. 
The supermolecular CMOs are also shown for reference. 
As can be seen, the canonical HOMOs are delocalized over both fragments, and localization (followed by re-canonicalization) helps to retain the delocalized nature of the $\pi$ and $\pi^{*}$ orbitals localized over each fragment. \\

Employing these localized orbitals, we carried out RILMO-TDA calculations for the uLE and uCT states. 
In order to demonstrate the flexibility of such an approach for this system, we also used Grimme's simplified TDA \cite{grimme2013simplified} approach (RILMO-simTDA) to calculate the uncoupled excitation energies and the electronic couplings given by Eq. (\ref{sTDA}).
The transitions considered in the uLE and uCT subspaces for the dimer consist of two  lowest $\pi \rightarrow \pi^{*}$ transitions of LE type for each of the chlorophyll, CLA1 and CLA2 (denoted as uLE$_{1}$ and uLE$_{2}$ respectively), and two lowest $\pi \rightarrow \pi^{*}$ of CT type -- CLA1 $\rightarrow$ CLA2 and CLA2 $\rightarrow$ CLA1 (denoted as uCT$_{1}$ and uCT$_{2}$ respectively). 
Table~\ref{uLE_cladimer_1} lists these energies for two of the conformations shown in Fig.~\ref{Structures}c, using both of these methods and PbE-sTDA. 
Although the full TDA and simplified TDA methods present some differences in the excitation energies, they both predict a distinct drop in the CT energies for conformation 2, although the latter yields a more drastic change probably owing to the approximations introduced in the calculation of the coupling elements of the $\mathbf{A}$ matrix for simplified TDA. 

The effect of the truncation of the size of the virtual space on the uLE and the uCT energies for both the conformations showed a similar trend as before, with the uLE energies displaying a gradual convergence behaviour with the increase in the number of virtual orbitals, and the uCT energies showing little to no change (see Fig.~S7 in SI). 
Thus, overall, the RILMO-TDA excitation energies are in reasonable agreement with those of PbE-sTDA. Table~\ref{couplings_cladimer_1} also lists the corresponding couplings between the uLE and uCT states calculated from RILMO-TDA and RILMO-simTDA approaches.
Although there are minor differences between these two methods, the couplings of RILMO-TDA are in excellent agreement with that of PbE-sTDA. 
Furthermore, the truncation of the size of the virtual space shows no discernible changes in the couplings (see Fig.~S8 in the SI).

\section{Conclusions}

In this work, we presented a simple one-step embedding approach for generating a set of re-canonicalized intrinsic localized molecular orbitals (RILMOs) for each fragment starting from a set of canonical molecular orbitals (CMOs). As an extension to the valence ILMOs discussed in our previous work \cite{senjean2021generalization}, these orbital sets include also the so-called hard virtual orbitals that are needed for a quantitative description of electronically excited states. While fully localized on individual fragments, RILMOs are delocalized within a fragment and can thus retain the intrinsic $\pi$ and $\pi^{*}$ character encountered in isolated conjugated systems. 
Since the ILMOs are orthogonal by construction, our procedure eliminates the need for the construction of any projection operators. Nevertheless, the orbitals are comparable to a set of polarized orthonormal orbitals arising from a top-down projection-based embedding procedure.
We furthermore outlined a TDDFT strategy for obtaining LE and CT states and investigated 
the effect of truncating the size of the virtual space in the localization procedure on the energies of these states.
While the diabatic LE energies displayed a relatively slow convergence behaviour upon increasing the total number of virtual orbitals in the localization step, the diabatic CT energies were found to converge very quickly.
The excitation energies and electronic couplings obtained in this approximate diabatic basis of the LE and CT states are in reasonable agreement with those from subsystem DFT using projection-based embedding.
Our observations are in accordance with previous studies that showed a similar dependence of the diabatic excitation energies and electronic couplings on the diabatization procedure employed. \cite{liu2015general,tolle2020electronic}. 
Nonetheless, we believe that our procedure will be useful due to its simple implementation and possibilities for easy extension to other excited state methods (DFTB, BSE, equation-of-motion coupled cluster, etc). 
The current scheme can for instance be used as an analysis toolkit for studying ET and EET processes in photo-induced systems.

\section*{Supplementary Material}

See the supplementary material for 1) effect of truncation of virtual space on the uLE energies of isolated $C_{2}H_{4}-C_{2}F_{4}$ and adenine-thymine dimers, 2) effect of truncation of virtual space on the electronic couplings of both of these dimers, 3) Convergence of the RILMO-TDA states in the CD and RD pathway
for a water-ammonia dimer, 4) electronic couplings and orbital energies for both of these dimers with varying distances and 5) effect of truncation of virtual space on the uLE and uCT
energies and electronic couplings of the chlorophyll dimer

\begin{acknowledgments}
The authors would like to thank Dr. Johannes T{\"o}lle for valuable discussions and Prof. Johannes Neugebauer for providing access to the ethylene-tetrafluoroethylene structure used in this work. S.S. and L.V. acknowledge support from NWO via the CSER program and for access to the National Computing Facilities.
\end{acknowledgments}

\section*{AUTHOR DECLARATIONS}

\subsection*{Conflict of Interest}
The authors have no conflicts to disclose.

\subsection*{Author Contributions}
\textbf{Souloke Sen}: Data curation (lead); Investigation (equal); Methodology (equal); Validation (equal); Writing - original draft (lead). \textbf{Bruno Senjean}: Methodology (equal); Validation (equal); Writing - review \& editing (equal). \textbf{Lucas Visscher}: Conceptualization (lead); Methodology (equal); Supervision (lead); Investigation (equal); Validation (equal); Writing – review \& editing (equal).

\section*{Data Availability}
The data used in this work are available upon reasonable request.

\nocite{bensberg2019automatic,dreuw2005single,bockers2018excitation,chulhai2016external,chulhai2015frozen,hirata1999configuration,stratmann1998efficient}

\section*{References}
\bibliography{main}
\end{document}


\clearpage

\section{Table of Contents:}
\begin{table}[h]
 \label{TOC}
  \begin{tabular}{cclcccccccc}
  & 1. &  Effect of truncating the virtual space on the uLE energies of & &\\
  & &  the isolated $C_{2}H_{4}-C_{2}F_{4}$ and adenine-thymine dimers & Pg. S3 &\\
   & & & & \\
   & 2. &  Effect of truncating the virtual space on electronic couplings of & &\\
   & & the $C_{2}H_{4}-C_{2}F_{4}$ and adenine-thymine dimers & Pg. S4 &\\
   & & & & \\
   & 3. & Convergence of the RILMO-TDA states in the CD and RD pathways & & \\
   & & for a water-ammonia dimer & Pg. S5 & \\
   & & & & \\
   & 4. & LE/LE and LE/CT electronic couplings for the $C_{2}H_{4}-C_{2}F_{4}$ dimer & Pg. S6 &\\
    & & & & \\
    & 5. & uLE and orbital energies for the $C_{2}H_{4}-C_{2}F_{4}$ dimer & Pg. S7 &\\
    & & & & \\
    & 6. & LE/LE and LE/CT electronic couplings for the adenine-thymine dimer & Pg. S8 &\\
    & & & & \\
    & 7. & Orbital energies for the adenine-thymine dimer & Pg. S9 &\\
    & & & & \\
    & 8. & Effect of truncating the virtual space on the uLE and uCT & &\\
   & & energies and electronic couplings of the chlorophyll dimer & Pg. S10 &\\
  \end{tabular}
\end{table}

\clearpage

\section{1. Effect of truncating the virtual space on the uLE energies of the isolated  $C_{2}H_{4}-C_{2}F_{4}$ and adenine-thymine dimers}

\begin{figure*}
\centering
\includegraphics[width=1.0\textwidth]{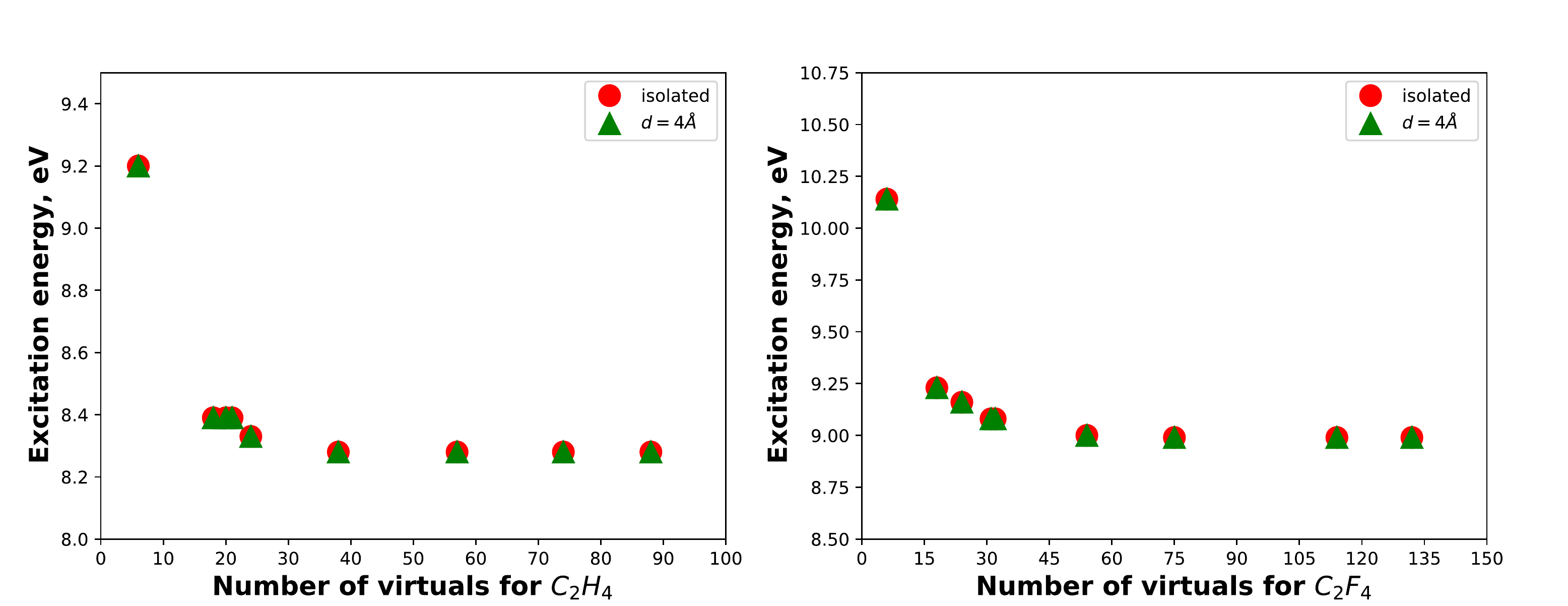}\\
\caption{Effect of truncation of the number of virtual orbitals on the uLE$_1$ (left panel) and uLE$_2$ (right panel) excitation energies for the isolated fragments and the $C_{2}H_{4}- C_{2}F_{4}$ dimer at $d=4~$\AA.} \label{E_L1_L2_ModifyVirt_C2h4c2f4}
\centering
\includegraphics[width=1.0\textwidth]{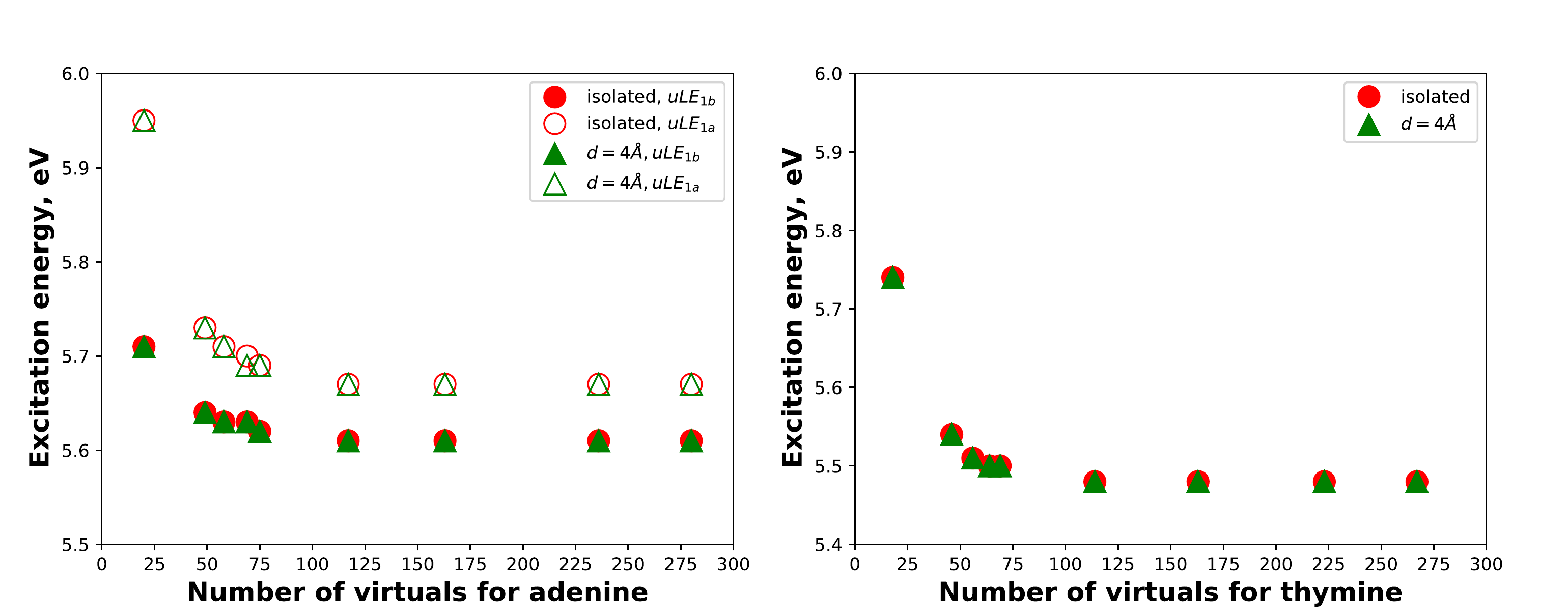}
\caption{Effect of truncation of the number of virtual orbitals on the uLE$_{1a}$, uLE$_{1b}$ (left panel) and uLE$_2$ (right panel) excitation energies for the isolated fragments and the adenine-thymine dimer at $d=4~$\AA.} \label{E1_E2_L1_L2_ModifyVirt_AT}
\end{figure*}

\clearpage

\section{2. Effect of truncating the virtual space on electronic couplings of 
the $C_{2}H_{4}-C_{2}F_{4}$ and adenine-thymine dimers}

\begin{figure}[h]
\centering
\includegraphics[width=1.0\textwidth]{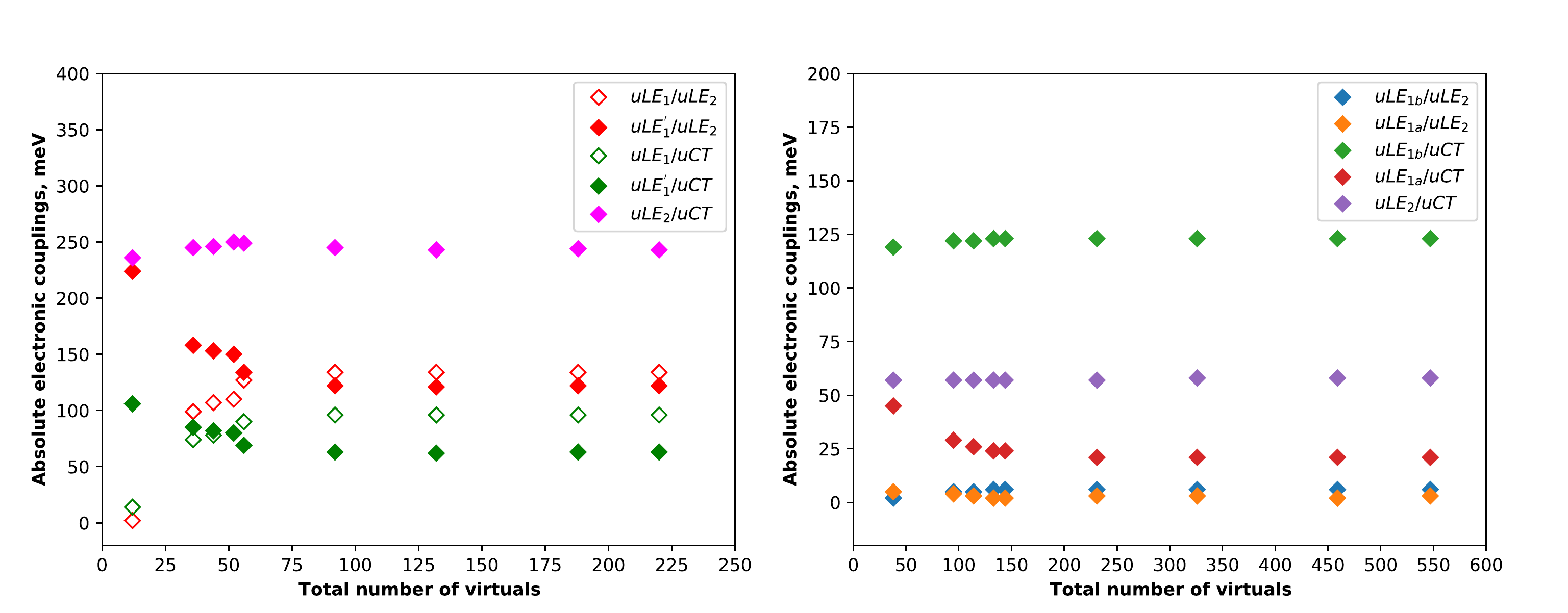}
\caption{Absolute electronic couplings between the uLE and uCT states of the $C_{2}H_{4}-C_{2}F_{4}$ (left panel) and adenine-thymine (right panel) dimers with respect to the number of hard virtual
orbitals used in the localization procedure, using the RILMO-TDA method.}
\label{Couplings_TrunVar_final_c2h4c2f4}
\end{figure}

\clearpage

\section{3. Convergence of the RILMO-TDA states in the CD and RD pathways for a water-ammonia dimer}

\begin{figure}[h]
\centering
\includegraphics[width=1.0\textwidth]{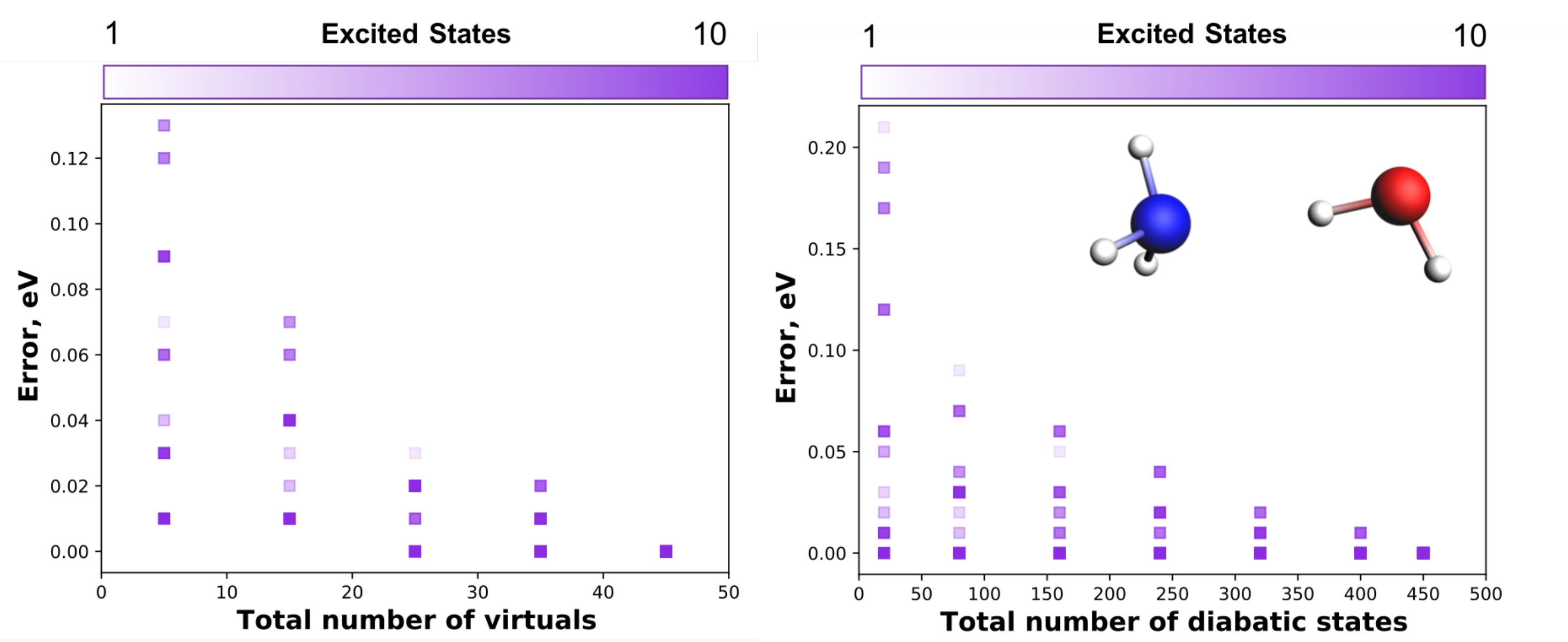}
\caption{Left panel: Error in excitation energy (as compared to the supermolecular CMO reference energies) due to RILMO truncation of
hard virtual orbitals for the lowest 10 RILMO-TDA (CD) states. Right panel: Error in excitation energy (as compared to the supermolecular CMO reference energies) due to truncation to $M$ diabatic states for the lowest 10 RILMO-TDA (RD,$M$) states, while keeping all the hard virtual orbitals in the localization procedure. Calculations were carried out at CAMB3LYP/DZP level of theory.}
\end{figure}

\clearpage

\section{4. LE/LE and LE/CT electronic couplings for the $C_{2}H_{4}-C_{2}F_{4}$ dimer}
\begin{figure}[h]
\centering
\includegraphics[width=0.9\textwidth]{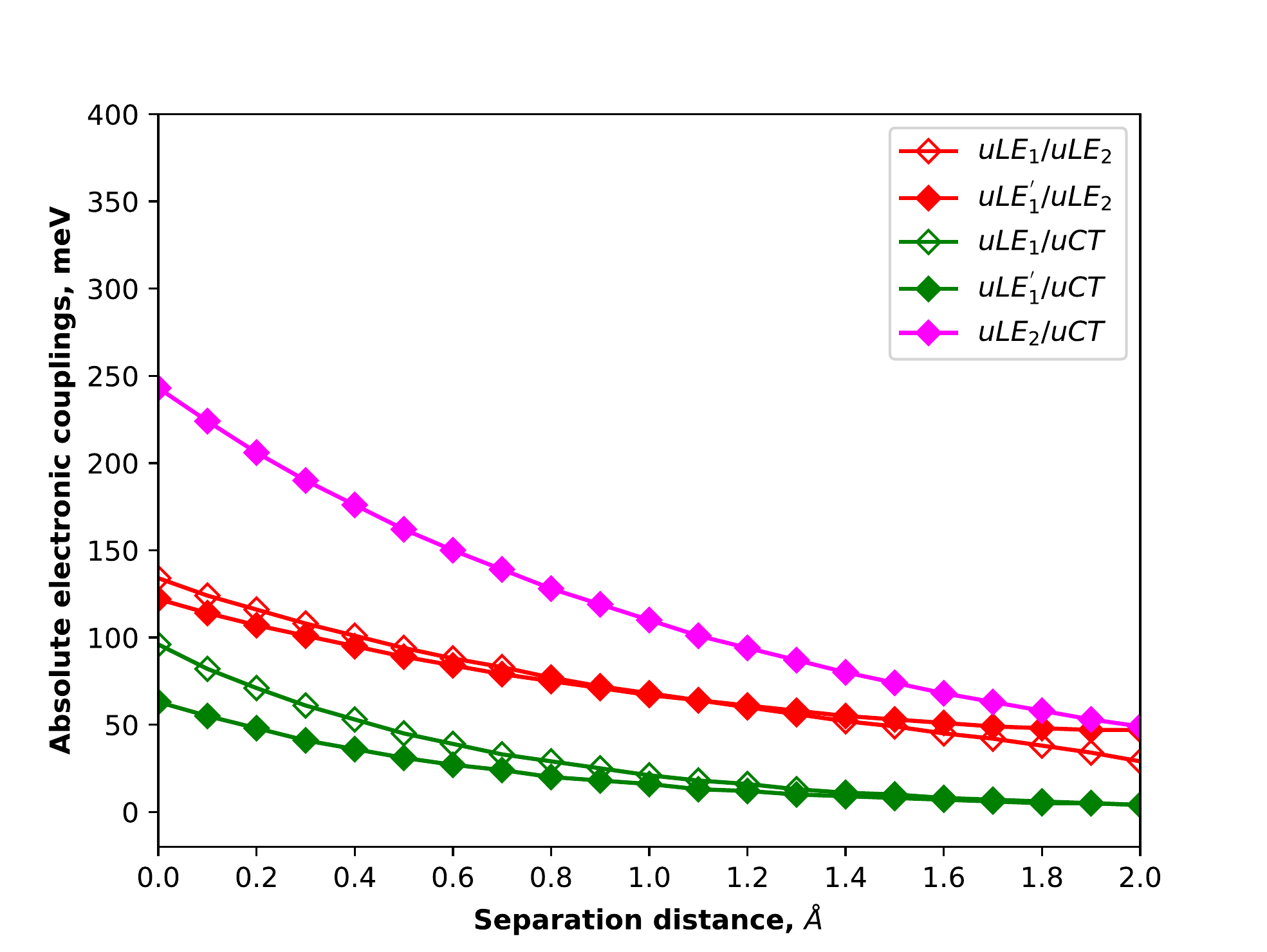}
\caption{Absolute electronic couplings between the uLE and uCT states of the $C_{2}H_{4}-C_{2}F_{4}$ dimer with respect to the separation distance, using the RILMO-TDA method.} 
\end{figure}

\clearpage

\section{5. uLE and orbital energies for the $C_{2}H_{4}-C_{2}F_{4}$ dimer}

\begin{table*}
  \caption{uLE energies of the $C_{2}H_{4} - C_{2}F_{4}$ dimer using RILMO-TDA and PbE-sTDA methods at varying distances, $\mathbf{d}$ (\AA). All units are in eV.}
 \label{uLE_C2h4c2f4_1}
  \begin{tabular}{ccccccccccc}
       & & &\vline & $\mathbf{d=0}$ & $\mathbf{d=0.5}$  &  $\mathbf{d=1.0}$ & $\mathbf{d=1.5}$ & $\mathbf{d=2.0}$  &\\ 
       \hline \\[-0.8em]
      &  &$\mathbf{uLE_{1}}$&\vline & 8.33 & 8.31 & 8.30 & 8.29  &  8.28 &\\
      &\textbf{RILMO-TDA}& $\mathbf{uLE_{1}^{'}}$ & \vline & 8.55 & 8.44 & 8.36 & 8.32 & 8.29 &\\
      &  & $\mathbf{uLE_{2}}$& \vline & 9.03 & 9.02   & 9.00  & 9.00  & 8.99 &\\   
     
  \end{tabular}
  \end{table*}

\begin{table*}
  \caption{Orbital energies associated with the uLE and uCT transitions of the $C_{2}H_{4} - C_{2}F_{4}$ dimer obtained from RILMO and PbE-sDFT methods at varying distances $\mathbf{d}$ (\AA). Also shown are the isolated fragment energies. The uCT and uLE states in the main text comprises of : 1) $uLE_{1} : HOMO(1) \rightarrow LUMO(1)$, 2) $uLE_{2} : HOMO(2) \rightarrow LUMO(2)/LUMO+1(2)$ and 3) $uCT : HOMO(2) \rightarrow LUMO(1)$. (1) and (2) refer to fragments $C_{2}H_{4}$ and $C_{2}F_{4}$ respectively. All units are in eV.}
 \resizebox{\columnwidth}{!}{%
  \begin{tabular}{ccccccccccc}
       & & &\vline & $\mathbf{d=0}$ & $\mathbf{d=0.5}$  &  $\mathbf{d=1.0}$ & $\mathbf{d=1.5}$ & $\mathbf{d=2.0}$  & $\mathbf{isolated}$ &\\ 
       \hline \\[-0.8em]
      &  &$\mathbf{LUMO(1)}$&\vline & 1.05 & 1.07 & 1.08 & 1.09  &  1.10 & 1.14 \\
      &  &$\mathbf{LUMO+1(2)}$&\vline & 1.69 & 1.65 & 1.63 & 1.61  &  1.60 & 1.56 \\      
      &\textbf{RILMO}&  & \vline & \\
      &  & $\mathbf{HOMO(1)}$& \vline & -9.42 & -9.37   & -9.34  & -9.32  & -9.31 & -9.27 &\\
      &  & $\mathbf{HOMO(2)}$& \vline & -9.15 & -9.18   & -9.20  & -9.22  & -9.23 & -9.27 &\\
     
     \hline \\[-0.8em]
     
      &  & $\mathbf{LUMO(1)}$&\vline  & 1.26 & 1.22  & 1.19  & 1.19  & 1.19 & 1.23 &\\
      &  & $\mathbf{LUMO(2)}$&\vline  & 1.84 & 1.80  & 1.76  & 1.74  & 1.72 & 1.68 &\\
      &\textbf{PbE-sDFT}&  &  \vline & \\
      &  & $\mathbf{HOMO(1)}$& \vline &  -9.37 &  -9.33   &  -9.31  & -9.29  & -9.28 & -9.24 &\\
      &  & $\mathbf{HOMO(2)}$& \vline &  -9.04 &  -9.07   &  -9.09  & -9.11  & -9.12 & -9.16 &\\
     
  \end{tabular}
  }
\end{table*}

\clearpage

\section{6. LE/LE and LE/CT electronic couplings for the adenine-thymine dimer}

\begin{figure}[h]
\centering
\includegraphics[width=1.1\textwidth]{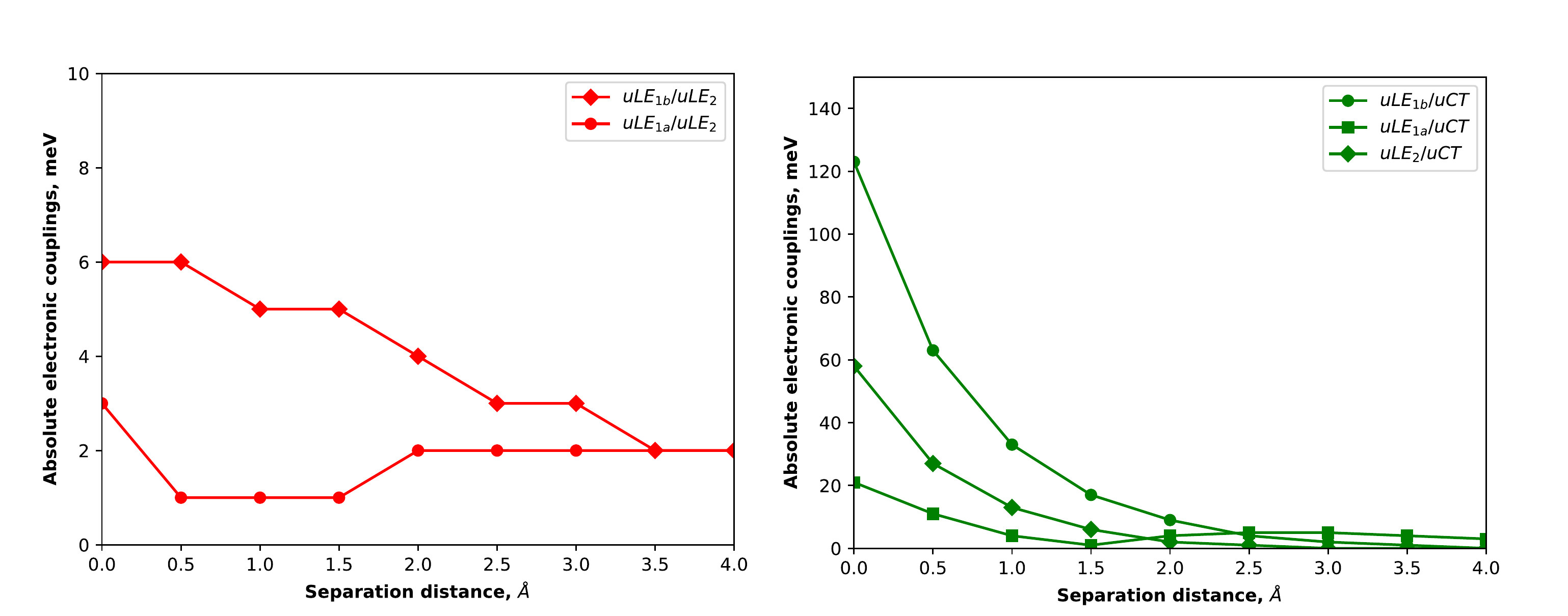}
\caption{Absolute electronic couplings between the uLE/uLE states (left panel) and the uLE/uCT states (right panel) of the adenine-thymine dimer with respect to the separation distance, using the RILMO-TDA method.} 
\end{figure}

\clearpage

\section{7. Orbital energies for the adenine-thymine dimer}

\begin{table*}
  \caption{Orbital energies associated with the uLE and uCT transitions of adenine-thymine dimer obtained from RILMO and PbE-sDFT methods at varying distances $\mathbf{d}$ (\AA). Also shown are the isolated fragment energies. The uCT and uLE states in the main text comprises of : 1) $uLE_{1a} : HOMO(1) \rightarrow LUMO(1)$, 2) $uLE_{1b} : HOMO(1) \rightarrow LUMO+1(1)/LUMO+2(1) $ , 3) $uLE_{2} : HOMO(2) \rightarrow LUMO(2)$ and $uCT : HOMO(1) \rightarrow LUMO(2)$ . (1) and (2) refer to fragments adenine and thymine respectively. All orbital energies are in eV.}
\resizebox{\columnwidth}{!}{%
  \begin{tabular}{ccccccccccc}
       & & &\vline & $\mathbf{d=0}$ & $\mathbf{d=1.0}$  &  $\mathbf{d=2.0}$ & $\mathbf{d=3.0}$ & $\mathbf{d=4.0}$  & $\mathbf{isolated}$ &\\ 
       \hline \\[-0.8em]
      &  &$\mathbf{LUMO+1(1)/LUMO+2(1)}$&\vline & 1.17 & 1.14 & 1.16 & 1.16  &  1.15 & 1.14 &\\
      &  &$\mathbf{LUMO(1)}$&\vline & 0.39 & 0.36 & 0.35 & 0.34  &  0.33 & 0.33 &\\
      &\textbf{RILMO}  & $\mathbf{LUMO(2)}$& \vline & -0.23 & -0.28  & -0.31  & -0.32  & -0.33 & -0.36 &\\ 
      &  & $\mathbf{HOMO(1)}$& \vline & -7.72 & -7.72   & -7.73  & -7.74  & -7.74 & -7.75 &\\
      &  & $\mathbf{HOMO(2)}$& \vline & -8.36 & -8.39   & -8.40  & -8.42  & -8.43 & -8.46 &\\
     
     \hline \\[-0.8em]
     
      &  &$\mathbf{LUMO+1(1)}$&\vline & 1.40 & 1.32 & 1.30 & 1.30  &  1.29 & 1.28 &\\
      &  &$\mathbf{LUMO(1)}$&\vline & 0.59 & 0.51 & 0.50 & 0.48  &  0.48 & 0.47 &\\
      & \textbf{PbE-sDFT} & $\mathbf{LUMO(2)}$& \vline & -0.04 & -0.15  & -0.19  & -0.20  & -0.21 & -0.25 &\\ 
      &  & $\mathbf{HOMO(1)}$& \vline & -7.60 & -7.60   & -7.60  & -7.61  & -7.61 & -7.63\\
      &  & $\mathbf{HOMO(2)}$& \vline & -8.26 & -8.29   & -8.32  & -8.33  & -8.34 & -8.38\\
  \end{tabular}
  }
\end{table*}

\clearpage

\section{8. Effect of truncating the virtual space on the uLE and uCT energies and electronic couplings of the chlorophyll dimer}

\begin{figure}[h]
\centering
\includegraphics[width=1.0\textwidth]{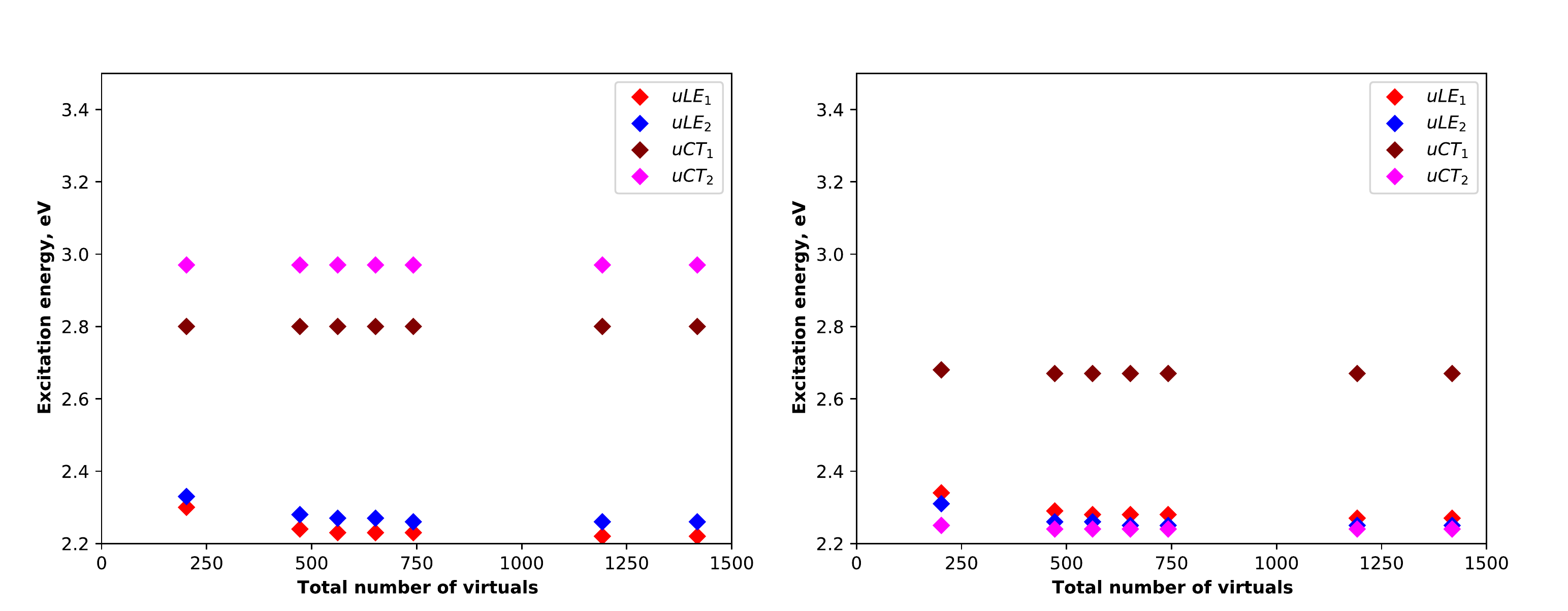}
\caption{Excitation energies of the uLE and uCT states of conformation 1 (left panel) and conformation 2 (right panel) with respect to the number of hard virtual orbitals used in the localization procedure, using the  RILMO-TDA method.} \label{Energies_conf1_conf2_TrunVar}
\end{figure}

\begin{figure}[h]
\centering
\includegraphics[width=1.0\textwidth]{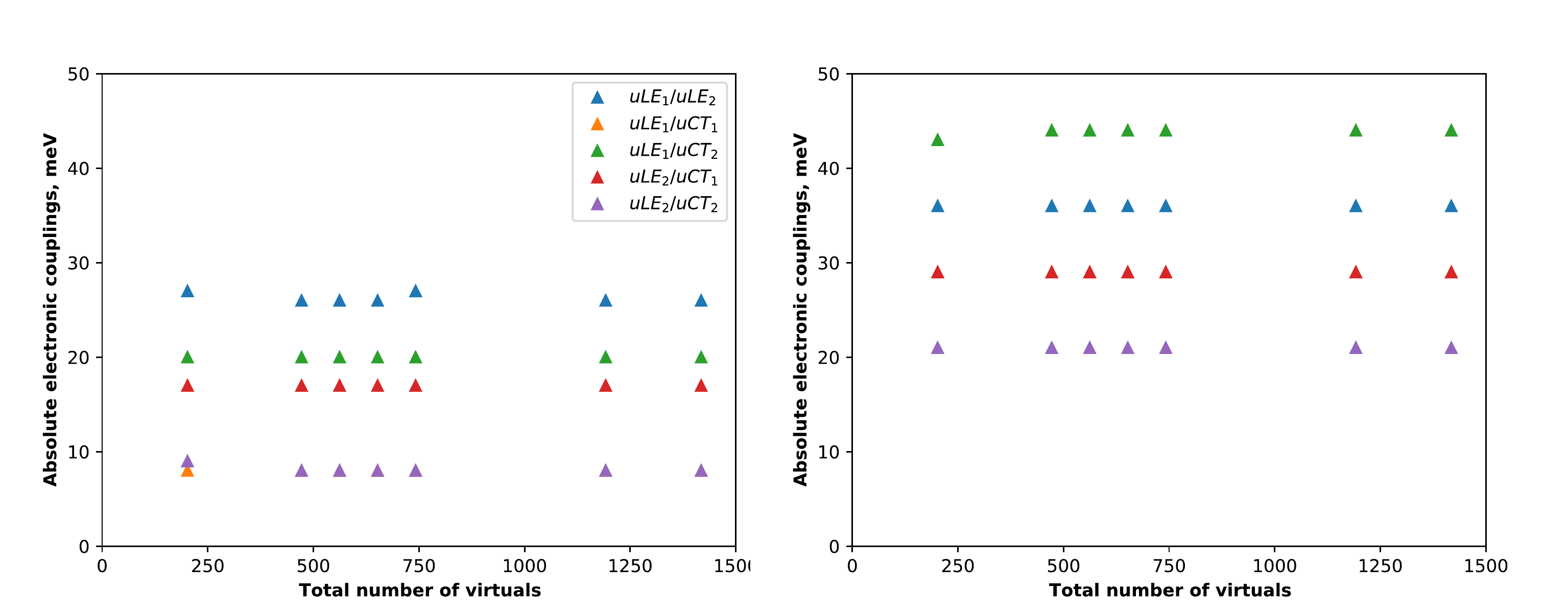}
\caption{Absolute electronic couplings between the uLE and uCT states of conformation 1 (left panel) and conformation 2 (right panel) with respect to the number of hard virtual orbitals used in the localization procedure, using the  RILMO-TDA method.} \label{Couplings_conf1_conf2_TrunVar}
\end{figure}

\clearpage